\documentclass[journal]{IEEEtran}
\usepackage{cite}
\usepackage{siunitx}
\usepackage[latin1]{inputenc}
\usepackage{color}
\usepackage{multirow}

\ifCLASSINFOpdf
   \usepackage[pdftex]{graphicx}
\else
\fi
\usepackage{amsmath}

\usepackage{stfloats}
\begin{document}
%
\title{Superconducting Wireless Power Transfer Beyond \mbox{5 kW} at High Power Density for Industrial Applications and Fast Battery Charging}
%
%
%

\author{Christoph~Utschick,
        Cem~Som,
        Ján~\v{S}ouc,
        Veit~Große,
        Fedor~Gömöry
        and Rudolf~Gross
\thanks{C. Utschick is with the Physik-Department, Technische Universität München, Munich, Germany and with Würth Elektronik eiSos GmbH \& Co. KG, Garching, Germany (email: christoph.utschick@tum.de).}
\thanks{C. Som is with Würth Elektronik eiSos GmbH \& Co. KG, Garching, Germany.}
\thanks{J. \v{S}ouc and F. Gömöry are with the Institute of Electrical Engineering, Slovak Academy of Sciences, Bratislava, Slovakia.}
\thanks{V. Große is with Theva Dünnschichttechnik GmbH, Ismaning, Germany.}
\thanks{R. Gross is with the Walther-Meißner-Institut, Bayerische Akademie der Wissenschaften, Munich, Germany and with the Physik-Department, Technische Universität München, Munich, Germany}
}

%
%

\markboth{}%
{Utschick \MakeLowercase{\textit{et al.}}: Superconducting Wireless Power Transfer Beyond \mbox{5 kW} at High Power Density for Industrial Applications and Fast Battery Charging}
%



\maketitle

\begin{abstract}
State of the Art Wireless Power Transfer (WPT) systems, based on conventional copper coils, are known to exhibit efficiencies well above 90\% when operated in the resonantly coupled mid-range regime. Besides full system efficiency, the area- and weight-related power densities of the transmission coils are key figures of merit for high power applications. This paper reports on a fully functional WPT system, consisting of single pancake  high-temperature superconducting (HTS) coils on the transmitter and the receiver side, which exceeds the power density of most conventional systems. Despite a compact coil size, a DC-to-DC efficiency above 97\% is achieved at 6 kW output power.  Next to the fundamental coil design, analytical and numerical simulations of the AC loss in the HTS coils are shown, taking into account both hysteresis and eddy current contributions. The results are validated by experimental AC loss measurements of single coils, obtained by a standard lock-in technique up to frequencies of 4 kHz. Finally, experimental results of the full system performance at different frequencies and load conditions are presented.
\end{abstract}


%
\IEEEpeerreviewmaketitle

\section{Introduction}
%
%
%
%
\IEEEPARstart{W}{ith} the rapid and continuously ongoing electrification of high power applications in the fields of industry, medicine and mobility, the need for convenient charging solutions has become more and more urgent. The technology of wireless power transfer (WPT) has emerged as one of the key enabling technologies to achieve a high degree of machine availability and to increase the general acceptance of electrical machinery on the market. In recent years, mature WPT solutions based on conventional technology have been successfully installed in electrical busses \cite{MomentumDynamics.2018} and ferries \cite{IPTTechnology.2019}. Charging powers of $\SI{200}{kW}$ have been demonstrated and fast charging systems with power levels up to $\SI{500}{kW}$ have been announced \cite{Wave.2019}. The transmission coils of these systems are well optimized and yield efficiencies above $\SI{90}{\percent}$. However, the area- and weight related power densities of conventional systems are limited by a Pareto front where a reduction of size or weight leads to a decrease in efficiency \cite{Bosshard.2015}. The available high power solutions are therefore only applicable to heavy machinery. Many mobile electrical machines like airborne applications, autonomous robots or race cars are, however, strongly restricted in weight and size and require alternative approaches.\\
Replacement of the conventional copper wire by a high-temperature superconducting (HTS) tape is discussed in literature as a promising approach to increase the efficiency and the transfer distance of WPT systems. A plethora of system designs have been proposed in \cite{XiaoYuanChenJianXunJin.2011,Kim.2012,Zuo.2015,Chen.2016,Narayanamoorthi.2019,Li.2019}, ranging from low- to high frequencies as well as from short- to long range systems with- and without relaying coils. Although the presented experimental results \cite{Zhang.2012,Yu.2015,Kang.2014,Inoue.2017b} suggest a remarkable performance of the HTS coils, the successfully transmitted power levels of less than $\SI{100}{W}$ are still disappointingly low. Furthermore, the AC losses in the HTS windings are not well discussed and the proposed coil designs are consequently not optimized. Measurements of the AC loss in typical HTS coils \cite{Pardo.2012b, Shen.2018} show that the loss per unit tape length exceeds the value of a single tape in its self-field by two orders of magnitude. A sophisticated coil design is therefore essential for an efficient WPT solution. Furthermore, the typical frequencies of WPT systems are beyond the frequency range where HTS tapes have been studied in literature and little knowledge about the high frequency behavior is available. Very recently, a few publications have studied the AC loss of HTS tapes in the kHz regime \cite{Shen.2017c, Inoue.2018, Zhou.2019c}. At such elevated frequencies eddy current losses in HTS tapes must be considered and are expected to even exceed the hysteresis loss.\\
In the following a fully functional WPT system, consisting of optimized single pancake HTS coils on transmitter and receiver side is presented that demonstrates the power transfer of $\SI{6}{kW}$ across an air gap of a few centimeters at an efficiency above $\SI{97}{\percent}$. In Section \ref{sec:fundamental-design} the fundamental system and coil design is presented and the interdependencies of the design parameters, including the choice of frequency and coil inductance are discussed. In Section \ref{sec:ac-loss-optimization}  the HTS coils are systematically optimized for low AC losses. Two calculation methods to determine the AC loss in the coil are introduced and compared. Hysteresis and eddy current losses in the low- and in the high frequency limits are considered and design guidelines for the inter-tape spacing, i.e. the distance between the individual turns are found. In Section \ref{sec:ac-loss-measurements}, the AC loss calculations are validated with experimental measurements based on a standard lock-in technique up to frequencies of $\SI{4}{kHz}$. Transport measurements of short tape samples and of the complete coils are shown. In Section \ref{sec:system-characterization}, a full characterization of the WPT system at different frequencies and load conditions is presented. Finally, it is shown that the reported measurement system allows for studying the AC loss of HTS coils in a frequency range from $\SI{80}{kHz}$ to $\SI{300}{kHz}$, which is not accessible by typical lock-in techniques. The obtained results allow to quantify the eddy current loss in the metallic layers of commercial HTS tapes and coils.


\section{Fundamental System and Coil Design}
\label{sec:fundamental-design}
\begin{figure}[!t]
	\centering
	\includegraphics[width=0.45\textwidth]{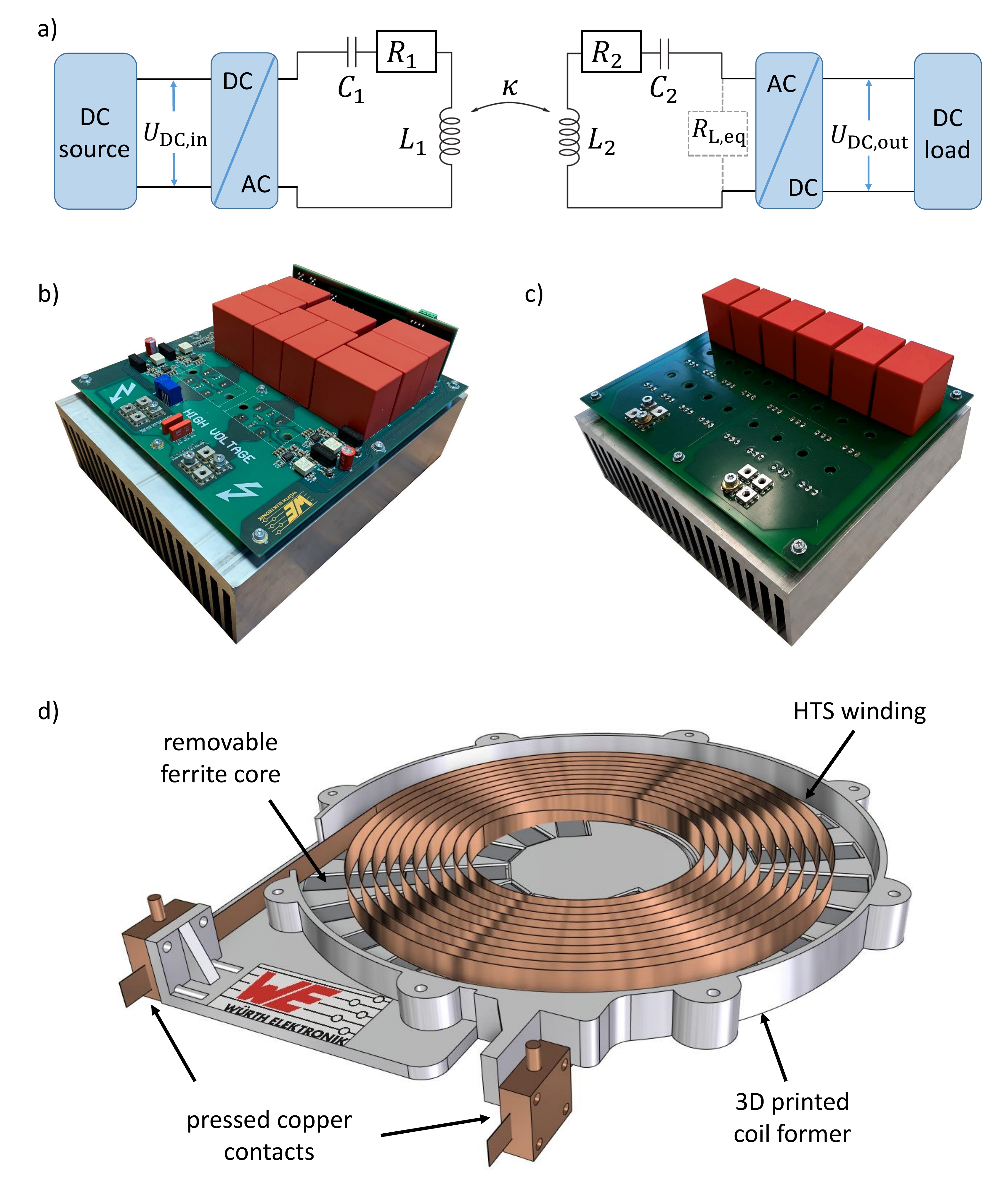}
	\caption{Overview of the fundamental system and coil design. In panel a), the complete power conversion chain of the inductively coupled magnetic resonant system is shown. In panel b) and c), pictures of the SiC based inverter and rectifier are shown. In panel d), the proposed coil design for superconducting WPT coils is shown. The coil consists of a distributed HTS winding and a removable ferromagnetic core which is mounted into a 3D printed coil former.}
	\label{fig:system-overview}
\end{figure}

The WPT system studied in this paper is an inductively coupled magnetic resonant circuit. An overview of the complete system, including all components of the power conversion chain, is depicted in Fig. \ref{fig:system-overview} a). The transmission coils, $L_1$ and $L_2$, with mutual inductance $M$, are connected to resonance capacitors, $C_1$ and $C_2$, to form a set of series-series compensated LC-resonators with identical resonance frequencies $f_\mathrm{r}$. The resonators are loosely coupled to each other with a coupling constant $\kappa=M/\sqrt{L_1 L_2}$. The inverter is based on a full-bridge design with two discrete MOSFET modules that are actively controlled by an external frequency generator with a duty cycle of 50\%. The inverter is driven at an operating frequency $f_\mathrm{op}$, which is slightly above $f_\mathrm{r}$. This ensures a positive phase angle of the input impedance and soft ZVS in the inverter. The rectifier is based on a passive diode bridge design, where each of the bridge elements is realized by 4  parallel Schottky diodes with two legs per diode. The strong parallelization results in small currents through the individual diode junctions and therefore low losses. Both devices, the inverter and the rectifier are based on silicon carbide (SiC) technology and can be used at frequencies of up to $\SI{1}{MHz}$ and power levels of up to $\SI{11}{kW}$. Pictures of both prototypes, which have been built by the company Würth Elektronik eiSos, are shown in Figs. \ref{fig:system-overview} b) and c). The transferred power is supplied by a DC source and is dissipated by a DC load with load resistance $R_\mathrm{L}$. The load is operated in a constant resistance mode and the transferred power is controlled by the input voltage on the transmitter side. The current and voltage levels at the input and at the output are measured with a high precision Yokogawa WT5000 power analyzer. The transmission coils are operated at 77 K and are cooled in a liquid nitrogen bath. The power electronics including the inverter, the rectifier and the resonant capacitors are operated at room temperature.\\
The characteristics of such a system are well understood and discussed in literature. Efficient power transfer generally requires two conditions. Firstly, the physically maximum possible coil-to-coil efficiency is limited by the quality factors $Q$ of the coils and the coupling constant $\kappa$ \cite{Bosshard.2015}
\begin{equation}
\eta_\mathrm{max}=\frac{\kappa^2 Q^2}{\left(1+\sqrt{1+\kappa^2 Q^2}\right)^2}\approx 1-\frac{2}{\kappa Q}\,.
\label{eq:coil-to-coil-efficiency}
\end{equation}
Secondly, an impedance matching between the input- and the load impedance is required, resulting in the condition \cite{Bosshard.2015}
\begin{equation}
R_\mathrm{L,eq}=\kappa\, \omega_\mathrm{r}L_2\,,
\label{eq:impedance-matching}
\end{equation}
where $R_\mathrm{L,eq}=8 R_\mathrm{L}/\pi^2$ is the equivalent load resistance in the resonant circuit and $\omega_\mathrm{r}=2\pi f_\mathrm{r}$ is the angular resonance frequency. Equation (\ref{eq:impedance-matching}) introduces a complex parameter interdependency between the coupling strength, the load resistance, the operating frequency and the coil inductance. Typically, $\kappa$ is limited by the spatial constraints of the application and $R_\mathrm{L}$ is defined by the desired current and voltage levels, leaving a design trade-off between the frequency and the coil inductance. Low frequency systems require a high coil inductance and vice-versa. When operated in a non-optimum working point, the system enters the over- or under-coupled regime where it exhibits frequency splitting or an unwanted voltage gain from source to load \cite{Wang.2004}.\\
\begin{table}[!t]
	\renewcommand{\arraystretch}{1.3}
	\caption{HTS tape properties}
	\label{tab:tape-parameters}
	\centering
	\begin{tabular}{lll}
		\hline
		Parameter& Tape A&Tape B\\
		\hline
		Manufacturer& Theva & Theva\\
		Product name& TPL 1100& TPL 4121\\
		Tape width&$\SI{12}{mm}$&$\SI{12}{mm}$\\
		Hastelloy substrate thickness&$\SI{100}{\micro\meter}$&$\SI{50}{\micro\meter}$\\
		HTS (GdBaCuO) thickness&$\SI{3}{\micro\meter}$&$\SI{3}{\micro\meter}$\\
		Silver overlayer&$\approx \SI{1.5}{\micro\meter}$ per side&$\approx \SI{1.5}{\micro\meter}$ per side\\
		Copper surround coating&--&$\SI{10}{\micro\meter}$ per side\\
		Min. bending diameter&$\SI{60}{mm}$&$\SI{40}{mm}$\\
		Critical current ($\SI{77}{K}$, s.f.)&$\SI{600}{A}$&$\SI{600}{A}$ \\
		\hline
	\end{tabular}
\end{table}
The fundamental coil design, proposed in this paper, is depicted in Fig. \ref{fig:system-overview} d). Following the design guidelines of conventional WPT coils \cite{Budhia.2011,Roshen.1988,Bosshard.2015}, the fabricated prototypes are realized as single pancake coils. The HTS winding is distributed across the available space with a precisely controlled inter-tape spacing between the individual turns. The winding is positioned on top of a removable ferromagnetic core (commercially available manganese-zinc (MnZn) ferrite), which is segmented into small stripes and arranged into a star-shaped structure. The stripes have dimensions of $\SI{45}{mm} \times \SI{13}{mm} \times \SI{2.5}{mm}$ and the ferrite volume is chosen large enough, such that the flux density inside the core at full power remains below the saturation field ($B_\mathrm{s}= \SI{820}{mT}$ at 77K). The current leads are positioned outside of the coil to avoid eddy current losses in the copper contacts. As HTS tapes do not allow to be bent or twisted on short length scales, the contacting of the innermost turn is challenging. In our design, the innermost turn is lowered, exits the coil former through a slit and is guided to the copper contact below the coil. The inter-tape spacing between the turns is realized by using a suitable spacer material (not shown in Fig. \ref{fig:system-overview} d)).\\
Two different types of HTS tapes have been used to build prototypes. In order to keep the eddy current losses low, a non-stabilized tape (type A) and a stabilized tape (type B) with a thin surround coating of $\SI{10}{\micro\meter}$ of copper on each side have been used in our experiments. The tapes, with a critical current of $\SI{600}{A}$ at $\SI{77}{K}$ in self-field, have been provided by the company Theva Dünnschichttechnik GmbH. All tape properties are listed in Tab. \ref{tab:tape-parameters}.

\section{AC loss optimization}
\label{sec:ac-loss-optimization}
\begin{figure}[!t]
	\centering
	\includegraphics[width=0.45\textwidth]{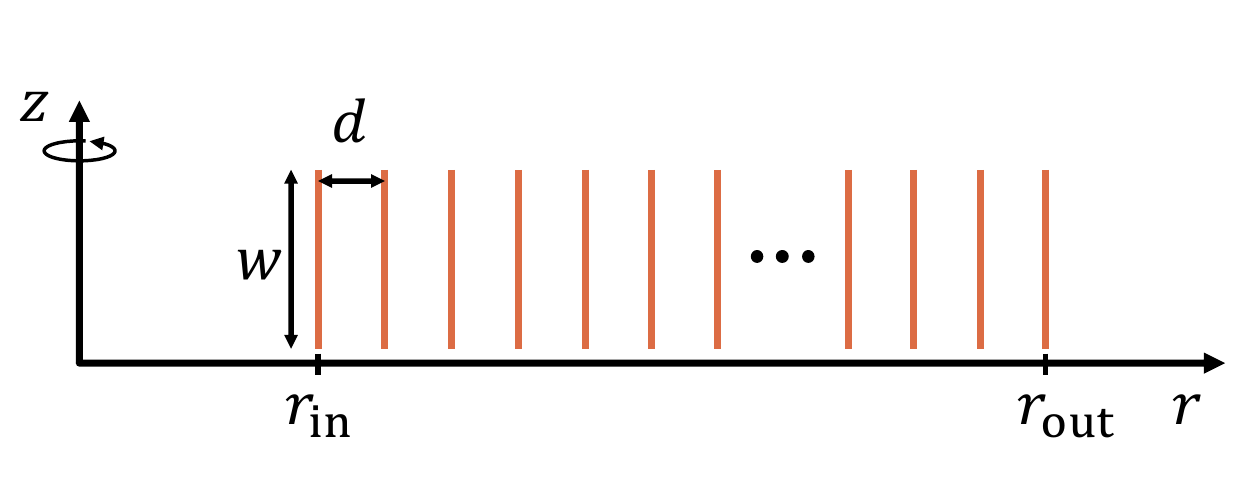}
	\caption{Simplification of the employed HTS winding to a rotational symmetric problem. The winding is parametrized by the inner coil radius $r_\mathrm{in}$, the number of turns $n$, the inter-tape spacing $d$ and the width of the tape $w$.}
	\label{fig:coil-simplification}
\end{figure}

For the calculation of the AC loss in the coil, the HTS winding is simplified to a rotational symmetric problem as depicted in Fig. \ref{fig:coil-simplification}. The winding is parametrized by the inner coil radius $r_\mathrm{in}$, the number of turns $n$, the inter-tape spacing $d$ and the width of the tape $w$. The fundamental design parameters to control the AC loss of the winding for a fixed number of turns are $d$ and $w$ \cite{Grilli.2014b}. In the following, the AC loss of an example coil with $n=10$ and $r_\mathrm{in}=\SI{5}{cm}$ is calculated and optimized. For simplification, the complex multilayer structure of commercial HTS tapes is reduced to a superconductor-metal (s-m) bilayer. The thickness of the metal layer is $t_\mathrm{m}$ and its material specific resistivity is $\rho$. For the case of a surround coating, where two identical metal layers are placed on the front- and on the backside of the tape, they are treated as a single layer with twice the thickness. Generally it is distinguished between the low and the high frequency regime. The transition frequency, where eddy currents in the metallic layer become comparable to the supercurrents in the HTS layer, depends on $t_\mathrm{m}$, $w$ and $\rho$ and is defined as \cite{Muller.1997}
\begin{equation}
f_\mathrm{tr}=\frac{\rho}{\mu_0 t_\mathrm{m} w}\,.
\label{eq:transition-frequency}
\end{equation}

\subsection{Simulation and Calculation Methods}
\label{sec:simulation-methods}
Although in literature \cite{Grilli.2014}, fully numerical simulations based on the H-formulation of the field quantities are considered as state of the art, here a fully analytical and a semi-analytical method are applied to calculate the AC losses.\\
In the fully analytical method the HTS winding is approximated as an infinite stack of tapes. For such a stack in its self-field, analytical equations for the hysteresis and eddy current losses have been derived in \cite{Muller.1999}. The loss per tape, per unit length, per cycle, in the low frequency limit, is given by
\begin{equation}
Q_\mathrm{hyst}= \frac{2\mu_0}{\pi^3}I_\mathrm{c}^2 \,F\left(\frac{d}{w},\frac{I_\mathrm{tr}}{I_\mathrm{c}}\right)\,,
\label{eq:z-stack-hyst}
\end{equation}
and
\begin{equation}
Q_\mathrm{eddy}= \frac{2\mu_0^2}{\pi^2}\frac{t_\mathrm{m}}{\rho}f I_\mathrm{c}^2 \,G\left(\frac{d}{w},\frac{I_\mathrm{tr}}{I_\mathrm{c}}\right)\,,
\label{eq:z-stack-eddy}
\end{equation}
where $I_\mathrm{tr}$ is the peak amplitude of the alternating transport current and $I_\mathrm{c}$ is the critical current of the tapes. The terms $F$ and $G$ are lengthy integral equations that are not reproduced here. Please note that the original equation of $Q_\mathrm{eddy}$ has been simplified for the special case that the widths of the superconducting strip and the metal strip in the s-m bilayer are identical.\\
It becomes obvious that both Eqs., (\ref{eq:z-stack-hyst}) and (\ref{eq:z-stack-eddy}), depend only on the geometry of the stack and the tape properties. The hysteresis loss per cycle is frequency independent and the eddy current loss per cycle scales with $f$. Due to the fact that the metal layers are strongly shielded by supercurrents, the hysteresis loss dominates the total loss of the stack in the low frequency limit. One can easily show that $Q_\mathrm{eddy}\ll Q_\mathrm{hyst}$ is valid for arbitrary stack geometries and arbitrary stabilizer thickness, as long as $f<f_\mathrm{tr}$ and $I_\mathrm{tr}<I_\mathrm{c}$. Due to symmetry reasons, the loss per unit length is identical in all the tapes of an infinite stack. For coil windings with sufficiently many turns, the boundary effects in the innermost and outermost turns can be neglected and Eq. (\ref{eq:z-stack-hyst}) is a good approximation of the average loss per unit length of the complete coil.\\ 
In the semi-analytical method on the other hand, the magnetic field distribution of the complete coil winding is computed with a numerical FEM simulation software (Ansys Maxwell). For the field computations in the low frequency limit, the HTS tape is modelled as a single thin strip that carries a uniform current density. Hereby it is assumed that the exact current distribution in the HTS tapes does not significantly influence the overall magnetic field distribution of the complete coil. Once the field distribution is known, the analytical Schönborg equation \cite{Schonborg.2001} is used to calculate the hysteresis loss for each individual turn, considering the transport current through the tape and the magnetic field generated by all the other turns of the coil. The loss per unit length in the n\textsuperscript{th} turn is given by
\begin{equation}
Q_\mathrm{n}=Q_\mathrm{Schoenborg}(I_\mathrm{tr},B_{\perp,\mathrm{n}})\,,
\end{equation}
where $B_{\perp,\mathrm{n}}$ is the perpendicular magnetic field component penetrating the n\textsuperscript{th} turn. As the amplitude and the direction of the magnetic field may vary locally along the tapes, $B_{\perp,\mathrm{n}}$ is determined by averaging the absolute value of $B_\perp$ along the width of each tape. It is noted explicitly that the self-field contribution of the turn under study must be subtracted from $B_{\perp,\mathrm{n}}$ as the input parameter for the Schönborg equation is only the externally applied field. The average hysteresis loss per unit length per cycle of the complete coil is given by 
\begin{equation}
Q_\mathrm{hyst}=\frac{\sum_{n=1}^{N}l_\mathrm{n}Q_\mathrm{n}}{\sum_{n=1}^{N}l_\mathrm{n}}\,,
\label{eq:hyst-loss-semi-analytical}
\end{equation}
where $N$ is the total number of turns and $l_\mathrm{n}$ is the length of the n\textsuperscript{th} turn.\\
As the strict assumption of uniform current density does not reproduce the exact field distribution in the direct vicinity of the tapes and neglects shielding currents between the turns, this method is expected to overestimate the loss in tightly wound coils ($d \rightarrow 0$). For HTS windings with sufficiently large inter-tape spacings however, it allows for a very fast and simple estimation of the AC loss. The assumption of uniform current density reduces the computation time for complex 3D structures to a few seconds.\\
Additionally, the numerical model was used to study the eddy current loss of the coil winding in the high frequency limit. In this regime, any effects related to superconductivity are insignificant. The total loss is dominated by eddy currents and the loss calculation becomes completely classical. In order to include the eddy current losses into the simulation, the metallic layer was modelled according to our bilayer model as a thin strip of thickness $t_\mathrm{m}$ and material specific resistance $\rho$. The transport current is uniform and flows entirely in the HTS layer. The eddy currents in each metal strip are driven by the self-field of the alternating transport current in the turn of study, and by the external field generated by all the other turns. The available calculation tools of the FEM software were used to determine the eddy current loss in each turn as a function of frequency for a fixed transport current amplitude. By changing the values of $t_\mathrm{m}$ and $\rho$, the influence of the different metallic layers was studied. In order to understand the high frequency behavior of the stabilized and non-stabilized tapes (compare Tab. \ref{tab:tape-parameters}), the AC loss in coils made of tapes with a $\SI{100}{\micro\meter}$ thick hastelloy substrate, a $\SI{3}{\micro\meter}$ thick silver layer and a $\SI{20}{\micro\meter}$ thick copper layer was simulated. The material specific resistances at $\SI{77}{K}$ are $\rho_\mathrm{hastelloy}^{77 K}$=$\SI{1240}{\nano\ohm \meter}$ \cite{Lu.2008}, $\rho_\mathrm{Ag}^{77 K}$=$\SI{2.7}{\nano\ohm \meter}$ \cite{Ekin.2007} and $\rho_\mathrm{Cu}^{77 K}$=$\SI{2}{\nano\ohm \meter}$ \cite{Ekin.2007}, respectively.\\
It is noted explicitly that the transport current in the metallic conduction path is zero in this model. A mechanism of current sharing between the superconducting and the metallic layers, as proposed in \cite{Zhou.2019c}, is not necessary to find quantitative agreement with experimental results.\\
%
\subsection{Simulation and Calculation Results}
\label{sec:simulation-results}
\begin{figure}[t]
	\centering
	\includegraphics[width=0.45\textwidth]{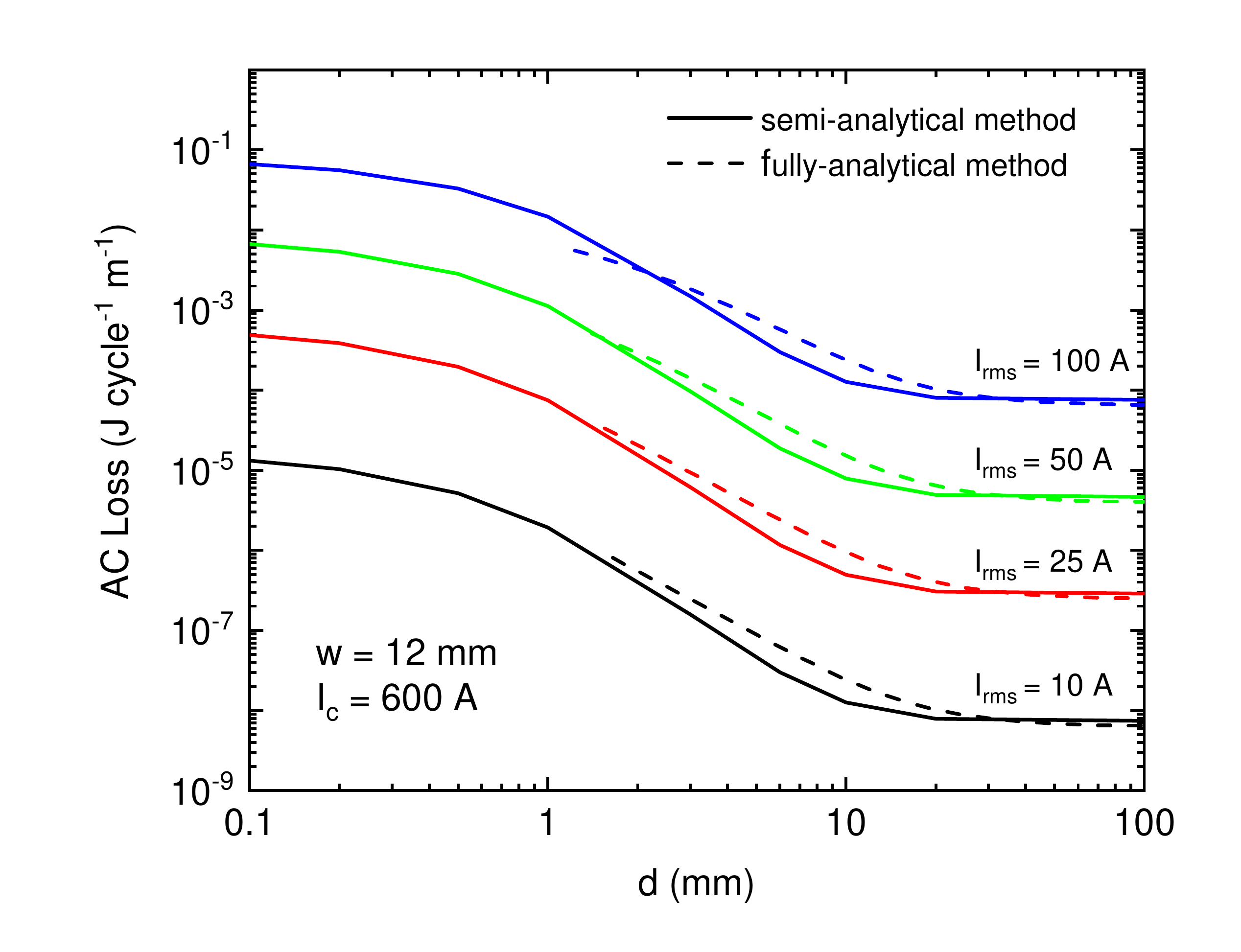}
	\caption{Calculated hysteresis loss of an HTS winding with $n=10$ in the low frequency limit. The loss per cycle per unit tape length is shown as a function of the inter-tape spacing $d$ for different current amplitudes. The fully-analytical (dashed lines, Eq. (\ref{eq:z-stack-hyst})) and the semi-analytical methods (solid lines, Eq. (\ref{eq:hyst-loss-semi-analytical})) are compared to each other.}
	\label{fig:ac-loss-comp-methods}
\end{figure}

\begin{figure}[!t]
	\centering
	\includegraphics[width=0.45\textwidth]{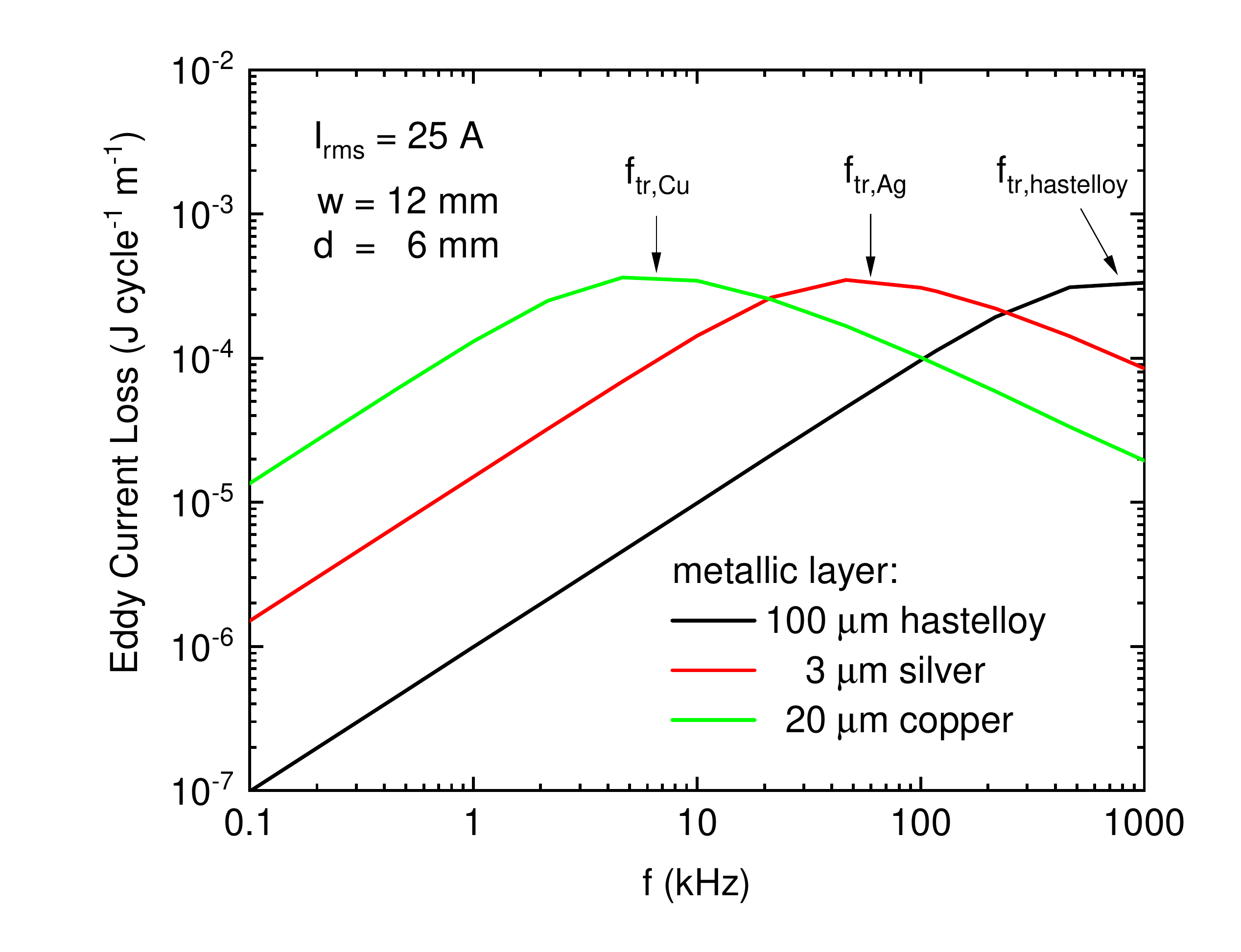}
	\caption{Simulated eddy current loss of an HTS winding with $n=10$, $w=\SI{12}{mm}$ and $d=w/2$ at a transport current of $I_\mathrm{rms}=\SI{25}{A}$ as a function of the frequency. The results of three typical metal layers are compared to each other. The individual transition frequencies of all layers are marked by the arrows. The transport current density in the HTS layer is assumed uniform and superconducting shielding currents are not considered. The material specific resistivities, that have been used to model the behavior of the metal strips at $\SI{77}{K}$, are defined in the main text.}
	\label{fig:eddy-current-loss}
\end{figure}

The results of the AC loss optimization are presented in Fig. \ref{fig:ac-loss-comp-methods}.
The hysteresis loss of the 10 turn HTS winding is shown as a function of $d$ for different current amplitudes in the low frequency limit. The width of the tape is fixed to $w=\SI{12}{mm}$ and the critical current is $I_\mathrm{c}=\SI{600}{A}$. The fully analytical and the semi-analytical calculation methods, i.e. Eqs. (\ref{eq:z-stack-hyst}) and (\ref{eq:hyst-loss-semi-analytical}), are compared to each other and show very good agreement. For large inter-tape spacings, the loss of the winding converges towards the value of a single straight tape in its self-field. With decreasing inter-tape spacing, the loss increases strongly. However, even a surprisingly small spacing is sufficient to keep the losses low. At a spacing of $d=w/2$, the loss of the winding is increased by a factor of 10, whereas at a spacing of $d=w$, the loss is only increased by a factor of 3, compared to the self-field case of a single tape. As the area of the coil increases with increasing $d$, the design of WPT coils results in a trade-off between minimizing the AC loss and maximizing the power per unit area of the coil.\\
When studying the influence of $w$ at a fixed value of $d$, one must take into account that a change of the width simultaneously changes the critical current of the tape. If the tape quality, i.e. the critical current density, is kept constant, then the AC loss of the winding decreases with increasing tape width and converges for $w>d$ into the limit of an equivalent infinite slab \cite{Clem.2007}.\\
The loss calculation suggests the following guidelines for optimized HTS windings:
\begin{itemize}
	\item The width of the tape should be chosen wide enough, such that the operating current is significantly below the critical current ($I_\mathrm{op}\ll I_\mathrm{c}$).
	\item The inter-tape spacing should be chosen in a range from $d=w/2$ to $d=w$. A further increase beyond $d=w$ brings little benefit as the winding approaches the single tape limit.
\end{itemize}
In the low frequency limit, the eddy current loss is much smaller than the hysteresis loss for all considered parameter variations and is therefore not shown in Fig. \ref{fig:ac-loss-comp-methods}.\\
In the high frequency regime on the other hand, the eddy current loss dominates the total loss of the tape. The analytical equations from \cite{Muller.1999} are not valid anymore and numerical simulations are required. In Fig. \ref{fig:eddy-current-loss}, the frequency dependence of the eddy current loss in the complete winding of the example coil with $w=\SI{12}{mm}$ and $d=w/2$ is presented at an exemplary transport current of $I_\mathrm{rms}=\SI{25 }{A}$.
The numerical simulation clearly shows the transition from the low into the high frequency regime for all three metal strips. The values of $f_\mathrm{tr}$ are marked with arrows in Fig. \ref{fig:eddy-current-loss}, and good agreement with the analytical expectation according to Eq. (\ref{eq:transition-frequency}) can be observed. In the low frequency regime, the simulation strongly overestimates the eddy current loss, as the superconducting shielding currents, which suppress the eddy currents are not considered. Beyond $f_\mathrm{tr}$, however, the skin depth becomes comparable to the thickness of the tape. Then, the shielding currents migrate into the metallic layer and the total loss of a m-s-strip becomes identical to the loss of a metal strip alone. Therefore, this method predicts the eddy current loss for $f>f_\mathrm{tr}$ correctly.  Fig. \ref{fig:eddy-current-loss} shows that the eddy current loss per cycle decreases with increasing frequency for $f>f_\mathrm{tr}$. As a result, the loss of a non-stabilized tape can be bigger than the loss of a stabilized one. Even the hastelloy substrate, which is typically completely neglected due to its high specific resistivity, can lead to significant loss beyond $\SI{100}{kHz}$.\\

\subsection{Fabricated Prototypes}
\label{sec:coil-prototype}
\begin{figure}[t]
	\centering
	\includegraphics[width=0.35\textwidth]{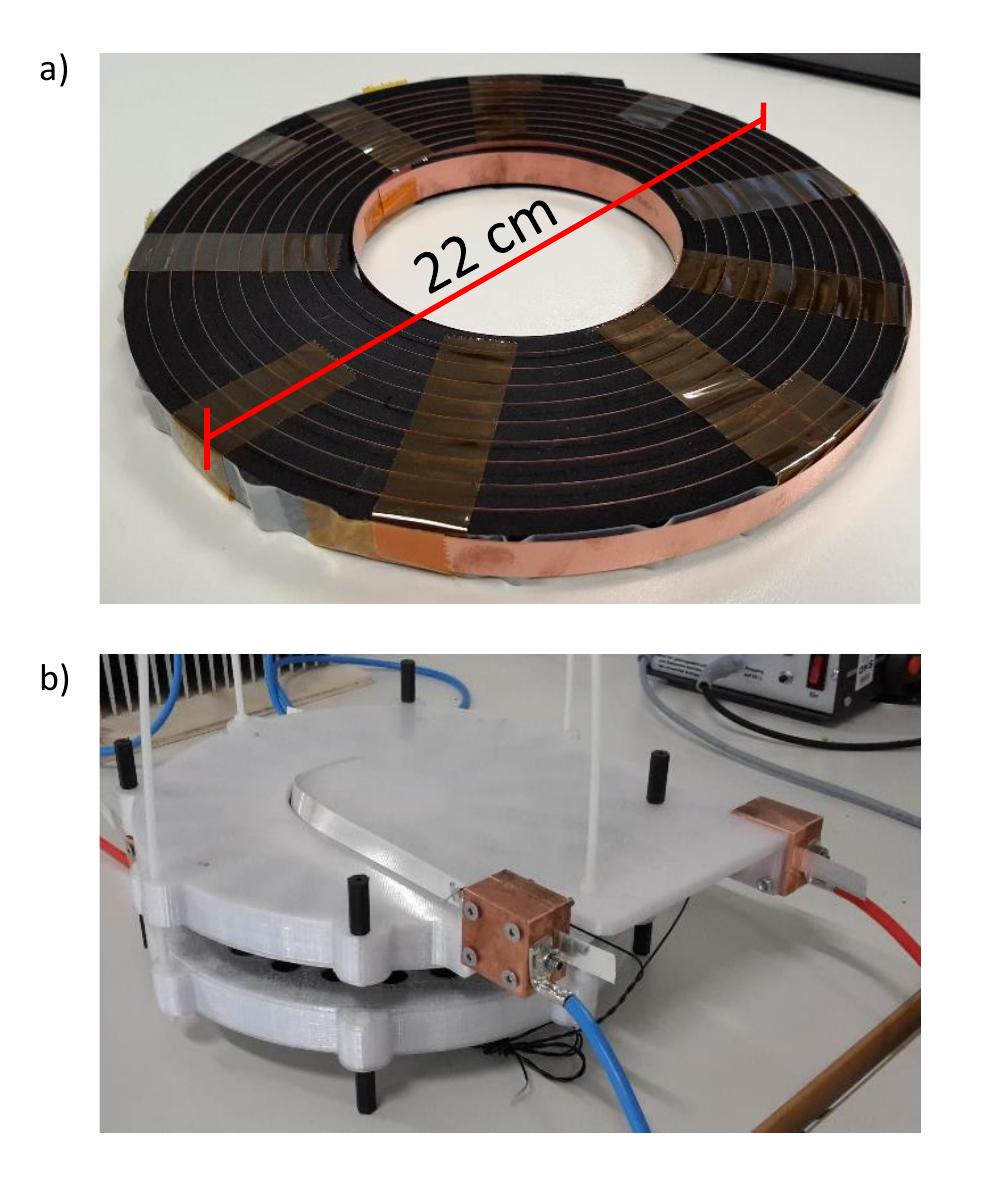}
	\caption{Pictures of the fabricated coil prototypes. In panel a), the HTS winding with spacer material between the individual turns is shown. In panel b) a pair of inductively coupled coils is shown. The receiver coil is mounted face down on top of the transmitter coil. Receiver and transmitter are spaced by an air gap of a few centimeters.}
	\label{fig:coil-prototype}
\end{figure}

Based on the performed loss optimization, two pairs of identical coil prototypes were realized with the parameters as listed in Tab. \ref{tab:coil-parameters}. One pair of coils is made of tape A, the other pair is made of tape B. Pictures of the HTS winding, before mounting it into the coil former, and of a coupled coil pair, as used for the WPT experiments are depicted in Fig. \ref{fig:coil-prototype} a) and b), respectively.\\
\begin{table}[t]
	\renewcommand{\arraystretch}{1.3}
	\caption{Properties of the realized prototype}
	\label{tab:coil-parameters}
	\centering
	\begin{tabular}{ll}
		\hline
		Parameter& Value\\
		\hline
		Inner coil radius&$\SI{5}{cm}$\\
		Outer coil radius&$\SI{11}{cm}$\\
		Number of turns&$10$\\
		Inter-tape spacing&$\SI{6}{mm}$\\
		Tape length&$\SI{5.4}{m}$\\
		\multirow{2}{*}{Inductance}&$\SI{20.5}{\micro\henry}$ (air coil)\\
		&\SI{24.4}{\micro\henry} (with magn. core @$\SI{77}{K}$)\\
		\hline
	\end{tabular}
\end{table}

\section{AC loss measurements}
\label{sec:ac-loss-measurements}
\begin{figure}[!t]
	\centering
	\includegraphics[width=0.45\textwidth]{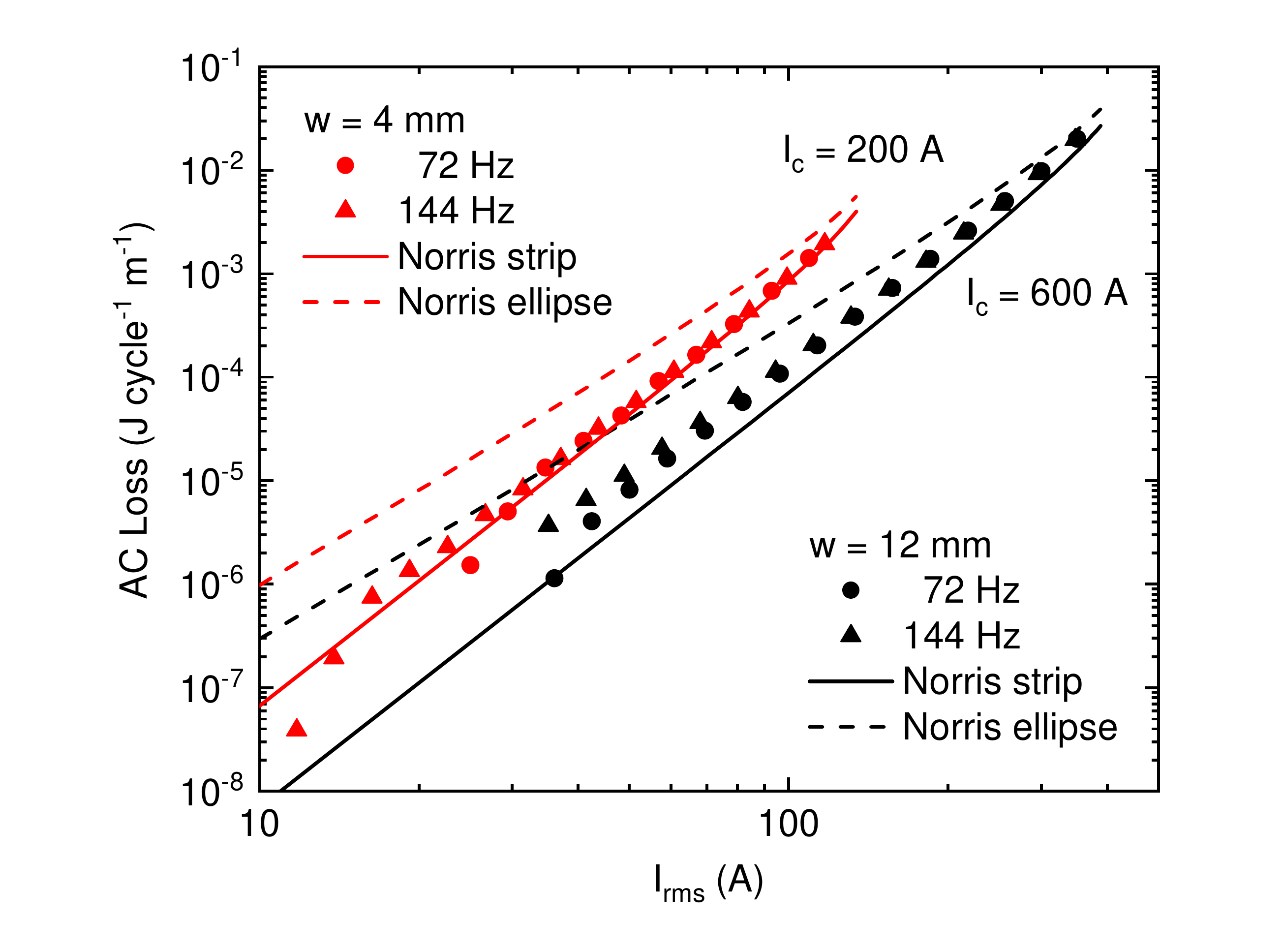}
	\caption{Transport loss measurement of short Theva tape samples with $w=\SI{4}{mm}$ and $w=\SI{12}{mm}$ at frequencies of $\SI{72}{Hz}$ and  $\SI{144}{Hz}$. The self-field loss is purely hysteretic and follows the analytical expectation. The measurement proves that the tilted ab-planes of the cuprate structure in Theva tapes do not influence the AC loss.}
	\label{fig:ac-loss-tape}
\end{figure}

\begin{figure}[!t]
	\centering
	\includegraphics[width=0.45\textwidth]{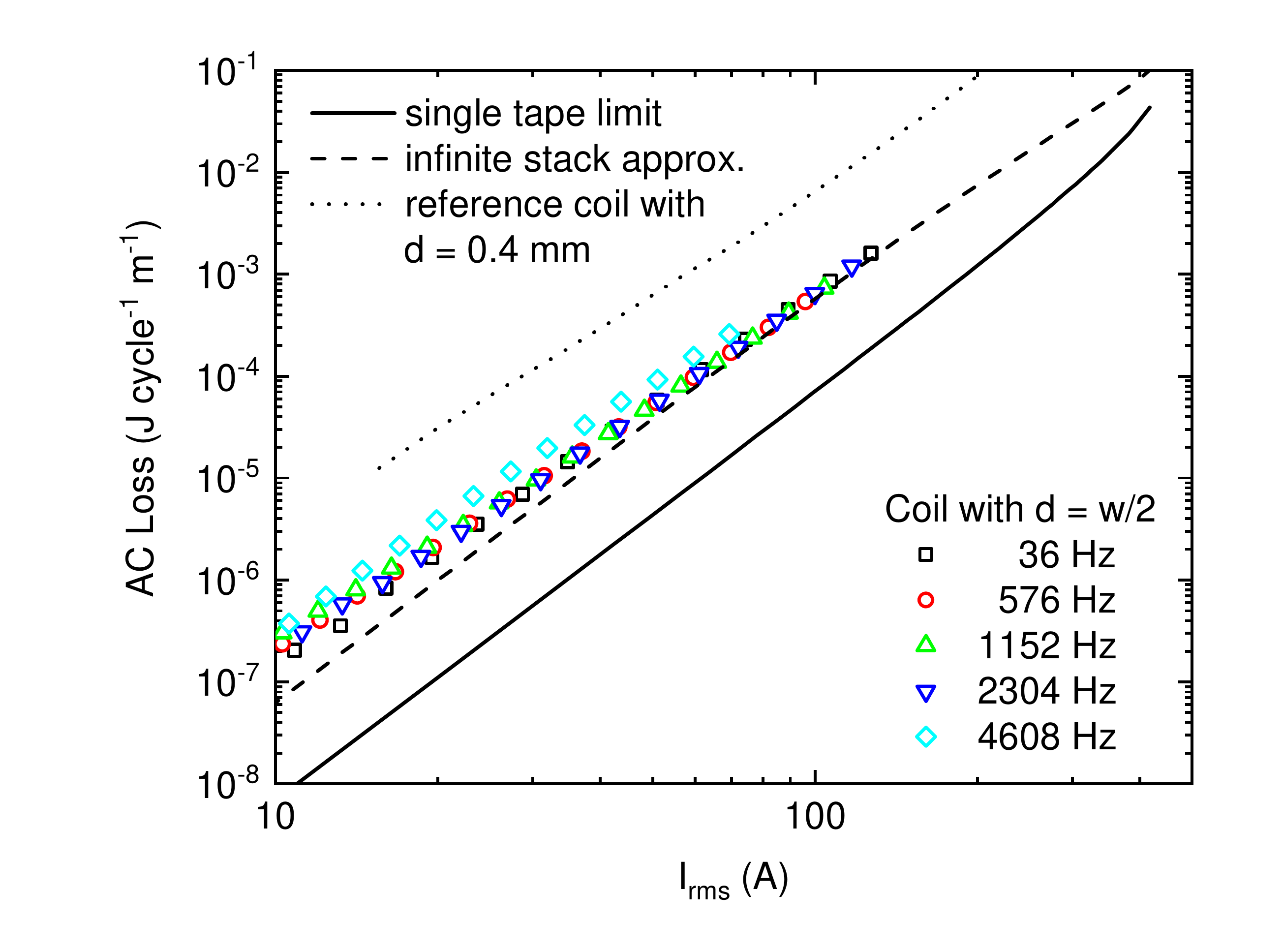}
	\caption{Transport loss measurement of the stabilized coil (tape B), with properties as defined in Tab. \ref{tab:coil-parameters}. The measured loss agrees very well with the infinite stack approximation (dashed line). In the calculation, $I_\mathrm{c}=\SI{600}{A}$ was assumed. The actual average critical current of the coil is $I_\mathrm{c}\approx\SI{550}{A}$. The limiting cases of a single tape in self-field (solid line) and of a tightly wound reference coil with $d=0.4$ (dotted line, data taken from \cite{Gomory.2013}) are shown as reference. The loss is frequency independent up to frequencies of $\SI{2.3}{kHz}$.  At $\SI{4.6}{kHz}$, the measured loss is slightly increased, compared to lower frequencies. This is most likely due to the onset of eddy current losses. Higher frequencies were not accessible by the available experimental setup.}
	\label{fig:ac-loss-coil}
\end{figure}

\begin{figure}[!t]
	\centering
	\includegraphics[width=0.45\textwidth]{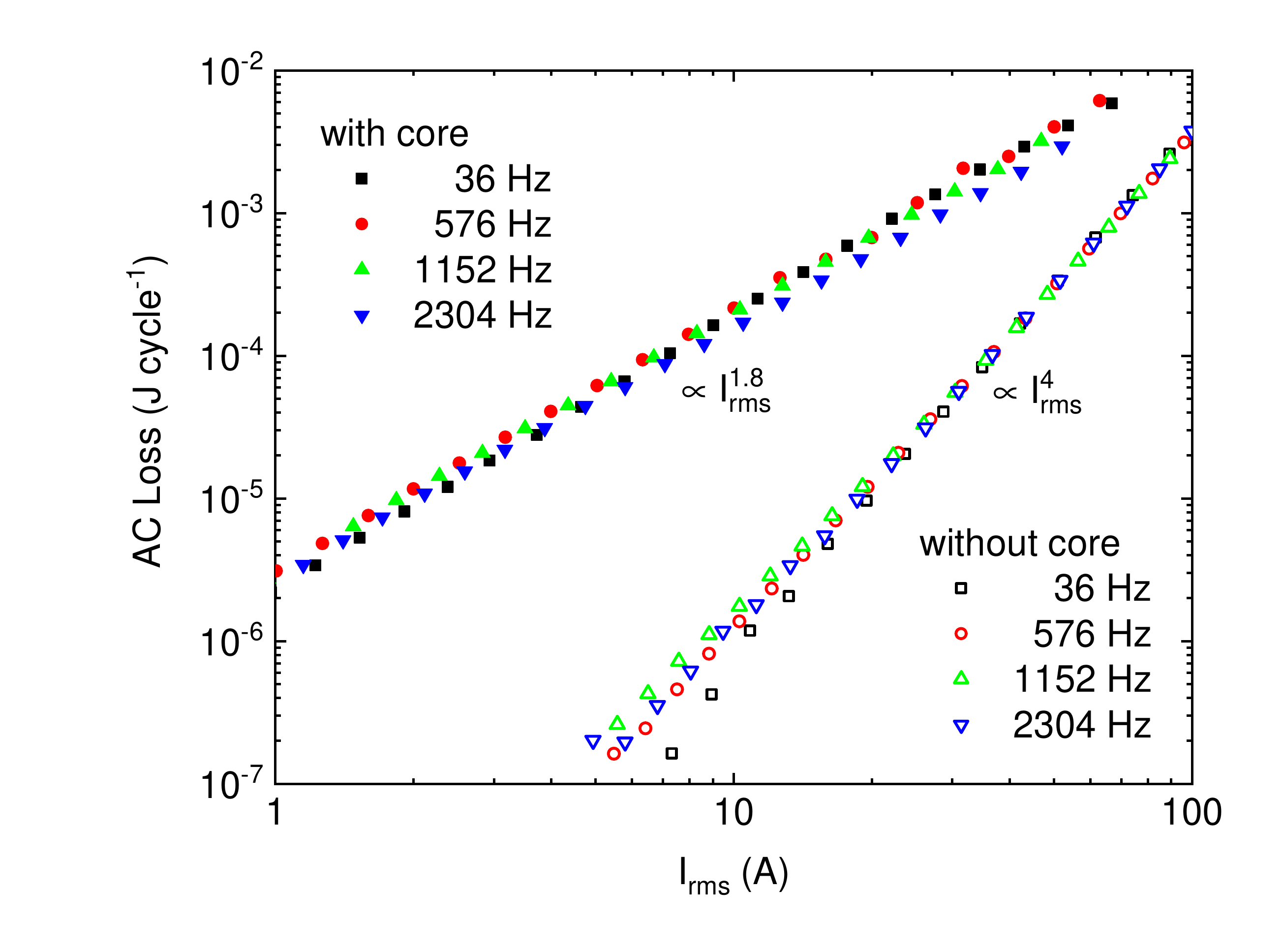}
	\caption{Influence of the ferromagnetic core on the total loss of the stabilized coil (tape B). Transport loss measurements, with and without the magnetic core, reveal that the magnetization loss in the cold core exceeds the loss of the HTS winding by several orders of magnitude. The magnetic loss is frequency independent and scales with $I^{1.8}$.}
	\label{fig:ac-loss-coil-with-ferrite}
\end{figure}

\begin{figure}[!t]
	\centering
	\includegraphics[width=0.45\textwidth]{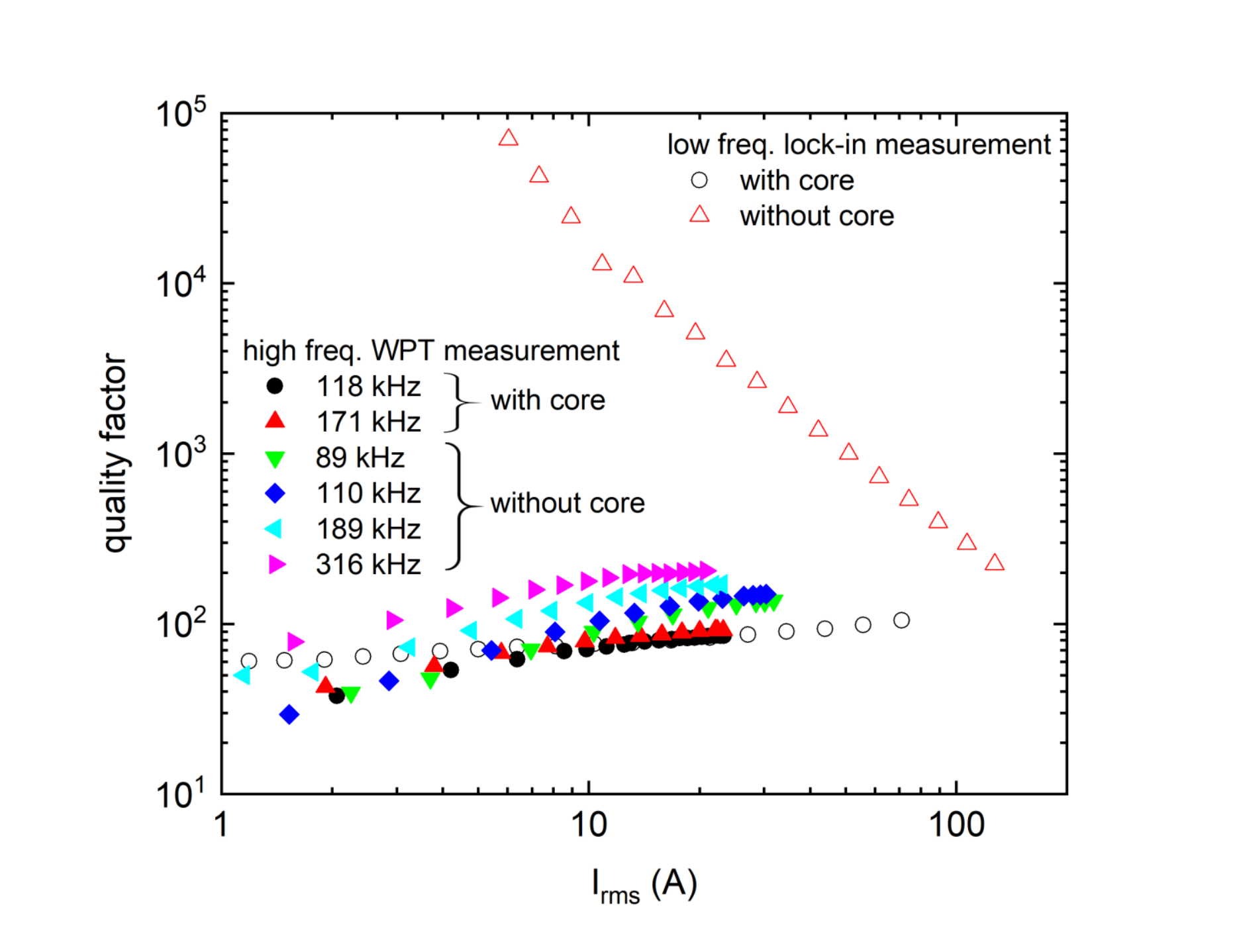}
	\caption{Measured quality factors of the HTS coils, with and without ferromagnetic core, in the low frequency limit as a function of the AC transport current. Without the core, the coil is dominated by the hysteresis loss in the winding. The Q-factor is frequency independent and scales with $1/I^2$. At low current amplitudes, the Q-factor exhibits values far beyond $10000$ and outperforms conventional copper coils. With the insertion of the cold ferromagnetic core, the supremacy of the superconducting winding is completely lost. For comparison also the quality factors in the high frequency regime are shown. They have been obtained by WPT experiments as explained in Sec. \ref{sec:system-characterization}.}
	\label{fig:q-factor}
\end{figure}

In order to validate the AC loss calculations, various measurements based on a standard lock-in technique, as described in \cite{Safran.2017}, were performed. Firstly the transport loss of short tape samples in self-field was measured. These measurements are particularly interesting, as the cuprate structure of the utilized Theva tapes differs from other manufacturers. They have the unique feature that the ab-planes of the HTS material are tilted with respect to the surface of the tape. The measurements of two different tapes with $w=\SI{4}{mm}$ and $w=\SI{12}{mm}$ are shown in Fig. \ref{fig:ac-loss-tape}. The measurement result are compared to the analytical expectation according to the Norris equations for strip and ellipse geometries \cite{Norris1970}. The expected loss is calculated for critical current values of $I_\mathrm{c}=\SI{200}{A}$ and $I_\mathrm{c}=\SI{600}{A}$. The actual critical currents of the tape samples, measured with a four probe method, are $I_\mathrm{c}=\SI{179}{A}$ and $I_\mathrm{c}=\SI{608}{A}$. These results prove that the self-field AC loss in Theva tapes follows the typical Norris equation.\\
Secondly, the transport loss of the fabricated coil prototypes was measured, with and without ferromagnetic core and at different frequencies up to $\SI{4.6}{kHz}$. The results of the stabilized coil (tape B), without core, are shown in Fig. \ref{fig:ac-loss-coil}. One can observe that the measured loss agrees nicely with the infinite stack approximation (dashed line). The reader is reminded that the only free parameter in Eq. (\ref{eq:z-stack-hyst}) is the critical current of the tape. In the calculation a value of $I_\mathrm{c}=\SI{600}{A}$ was assumed. According to the tapestar measurement, the actual average critical current of the utilized tape segment was $I_\mathrm{c}\approx \SI{550}{A}$. In order to avoid a possible damage of the coil, the critical current of the coil was not measured after the winding process. Due to the large spacing between the turns, however, it is expected that the $I_\mathrm{c}$ of the coil is not reduced compared to the tape in its self-field. The good agreement between calculation and measurement in Fig. \ref{fig:ac-loss-coil} has been obtained without a single fit parameter. As reference, the AC loss of the limiting cases of a single straight tape in self-field (solid line) and a tightly wound coil with $d=\SI{0.4}{mm}$ (dotted line, measurement data taken from \cite{Gomory.2013}) are also shown. Up to frequencies of $\SI{2.3}{kHz}$, the measured loss is frequency independent, i.e. of purely hysteretic nature. Evidently, at the highest frequency of $\SI{4.6}{kHz}$, the loss is slightly increased. As the transition frequency of the copper stabilized tape is at $f_\mathrm{tr}=\SI{6.6}{kHz}$, the increased AC loss could possibly be explained with the onset of eddy current losses. Repetition of the experiment on a non-stabilized coil (tape A), which has a higher transition frequency, showed indeed no increased loss at $\SI{4.6}{kHz}$. This allows to rule out the possibility of a systematic error. A further increase of the measurement frequency was not possible due to technical restrictions in the available experimental setup.\\
AC loss measurements of the coils with ferromagnetic core, revealed that the magnetization loss in the core exceeds the loss of the HTS winding by several orders of magnitude. A comparison of the measurements with and without core is shown in Fig. \ref{fig:ac-loss-coil-with-ferrite}. One can see that the magnetic core loss is frequency independent and scales with $I^{1.8}$. A detailed discussion of the magnetic core loss would exceed the scope of this work. However, it is important to mention that additionally the loss of a single core segment in external field was measured with the calibration free setup from \cite{Souc.2005}. The measurements have shown that the loss of a cold core at $\SI{77}{K}$ is increased by more than a factor of 10 compared to the loss at room temperature. Our magnetic loss measurements agree quantitatively with the results published in \cite{Claassen.2005}.\\
Finally, the quality-factor (Q-factor) of the coils was determined, which is defined as
\begin{equation}
\label{eq:quality-factor}
Q=2\pi \times \frac{\mathrm{stored\,energy}}{\mathrm{energy\,loss\,per\,cycle}}\,.
\end{equation}
The stored energy in the coil, is given by $LI_\mathrm{peak}^2/2=LI_\mathrm{rms}^2$ and the energy loss per cycle has been determined experimentally in Fig. \ref{fig:ac-loss-coil-with-ferrite}. The Q-factors of the fabricated coil, with and without ferromagnetic core, are shown as a function of the AC transport current in the low frequency limit in Fig. \ref{fig:q-factor}.
The Q-factor of the HTS winding is frequency independent, scales with $1/I^2$ and exhibits values far beyond $10000$ at low current amplitudes. Comparison to conventional copper coils, which hardly exceed Q-factors of $1000$, reveals that the superconducting hysteresis loss is no show-stopper for WPT coils. As long as the coil design follows the guidelines, the current amplitudes remain below $I_\mathrm{rms}\approx \SI{40}{A}$ and the operating frequency remains below $f_\mathrm{tr}$, an outstanding performance of the WPT system is to be expected. When a cold ferromagnetic core is inserted into the coil, the additional losses reduce the Q-factor to values below $100$ and the supremacy of the superconducting winding over conventional copper coils is completely lost. The best option would be to avoid the magnetic core at all. In case it is needed for shielding purposes, a sophisticated cryostat design with a warm core will be required. For comparison also the quality factors in the high frequency regime are shown. They have been obtained by WPT experiments as explained in Sec. \ref{sec:system-characterization}. The quality factors in the high frequency regime are dominated by eddy current loss and are therefore strongly reduced compared to the low frequency regime. It becomes obvious that the benefits of superconducting transmission coils can only be fully exploited in the low frequency regime.\\

\section{Full System Characterization}
\label{sec:system-characterization}
\begin{figure}[!t]
	\centering
	\includegraphics[width=0.45\textwidth]{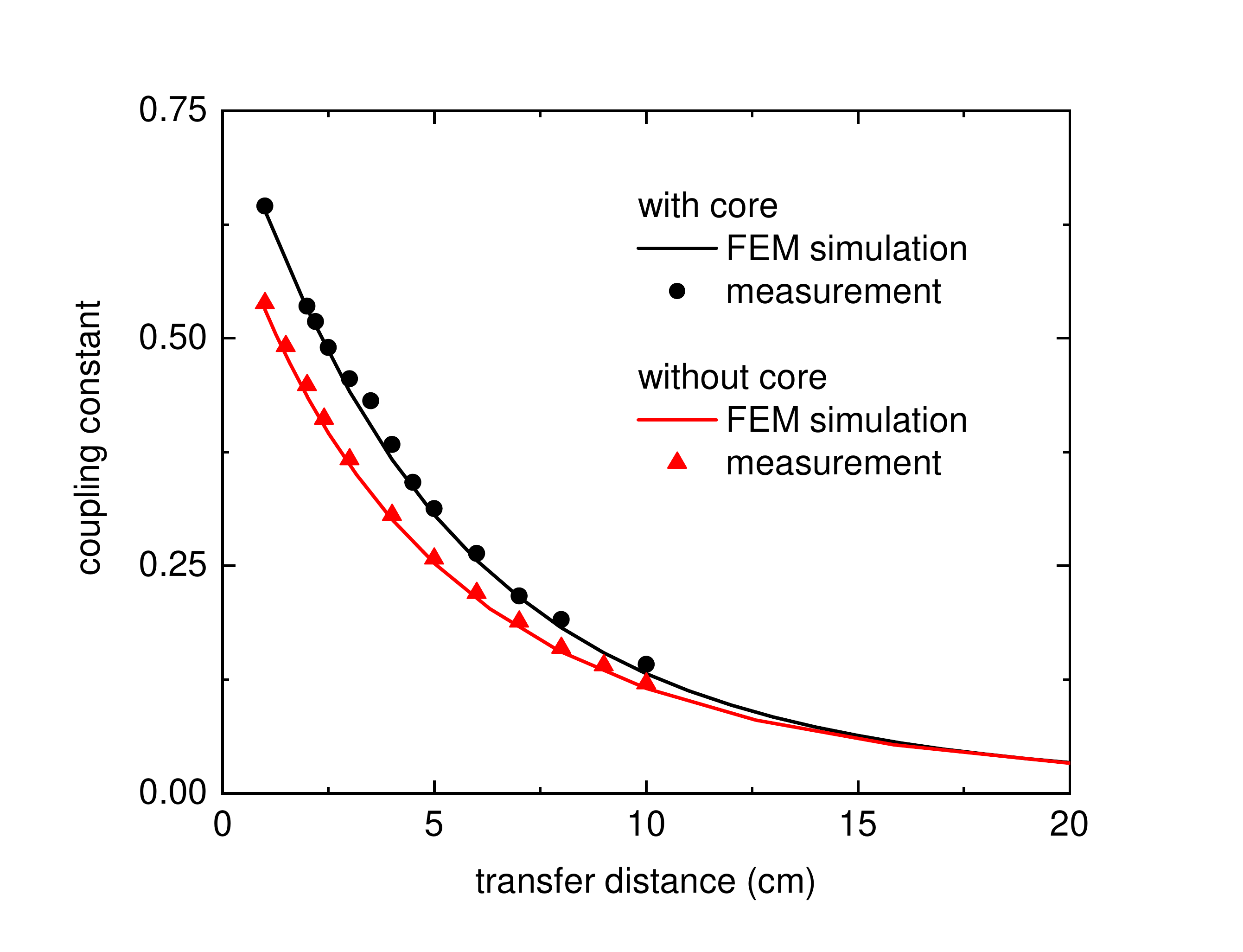}
	\caption{Measured and simulated coupling constant between transmitter and receiver coil as a function of the transfer distance. The cases with and without magnetic core are compared to each other.}
	\label{fig:coupling}
\end{figure}

\begin{figure}[!t]
	\centering
	\includegraphics[width=0.4\textwidth]{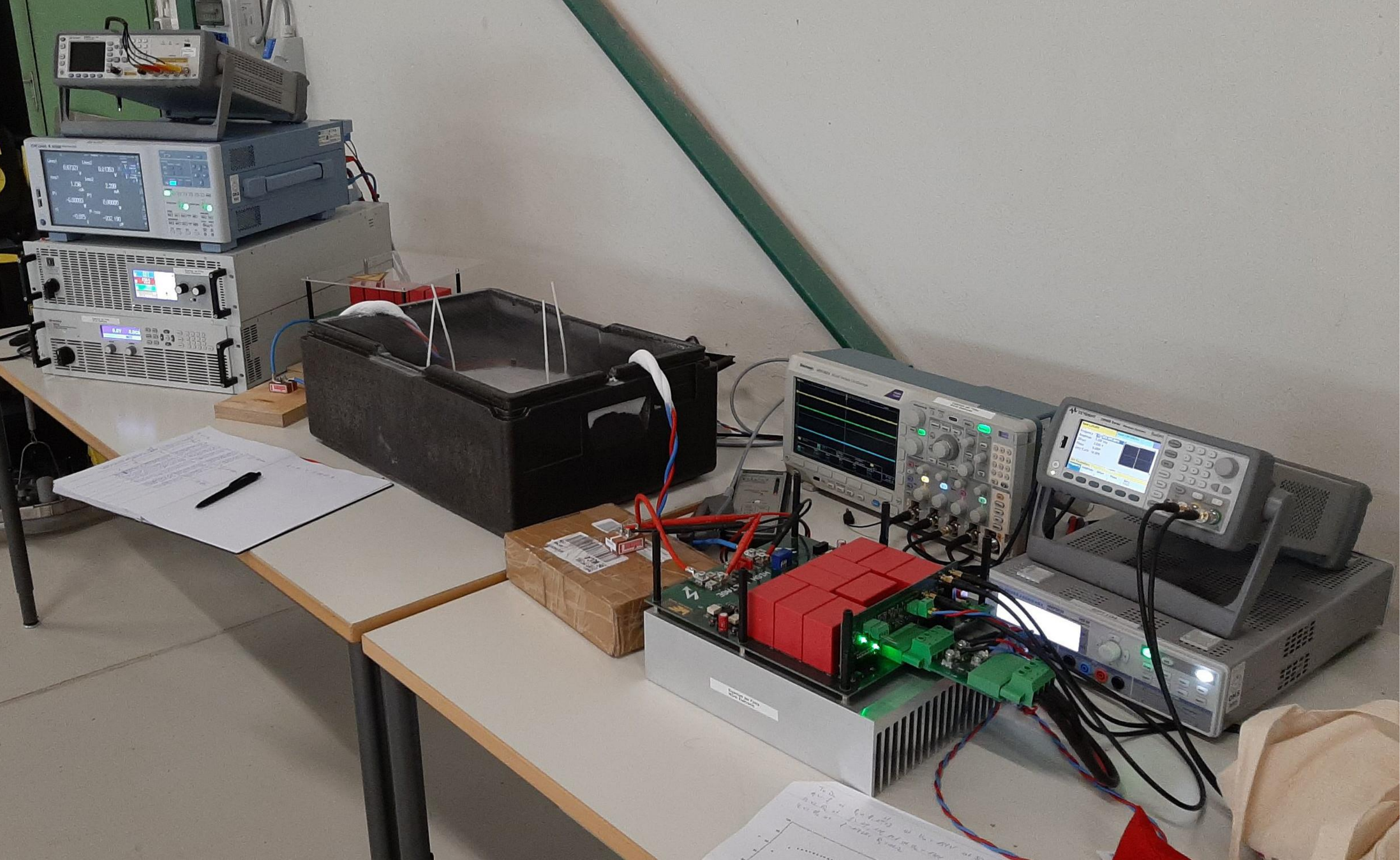}
	\caption{Photograph of the experimental setup including all measurement devices that are required for the full system characterization. The superconducting coils are inside the black box which is filled with liquid nitrogen. The rectifier is placed behind the black box and is not visible.}
	\label{fig:full-system}
\end{figure}

\begin{figure}[!t]
	\centering
	\includegraphics[width=0.45\textwidth]{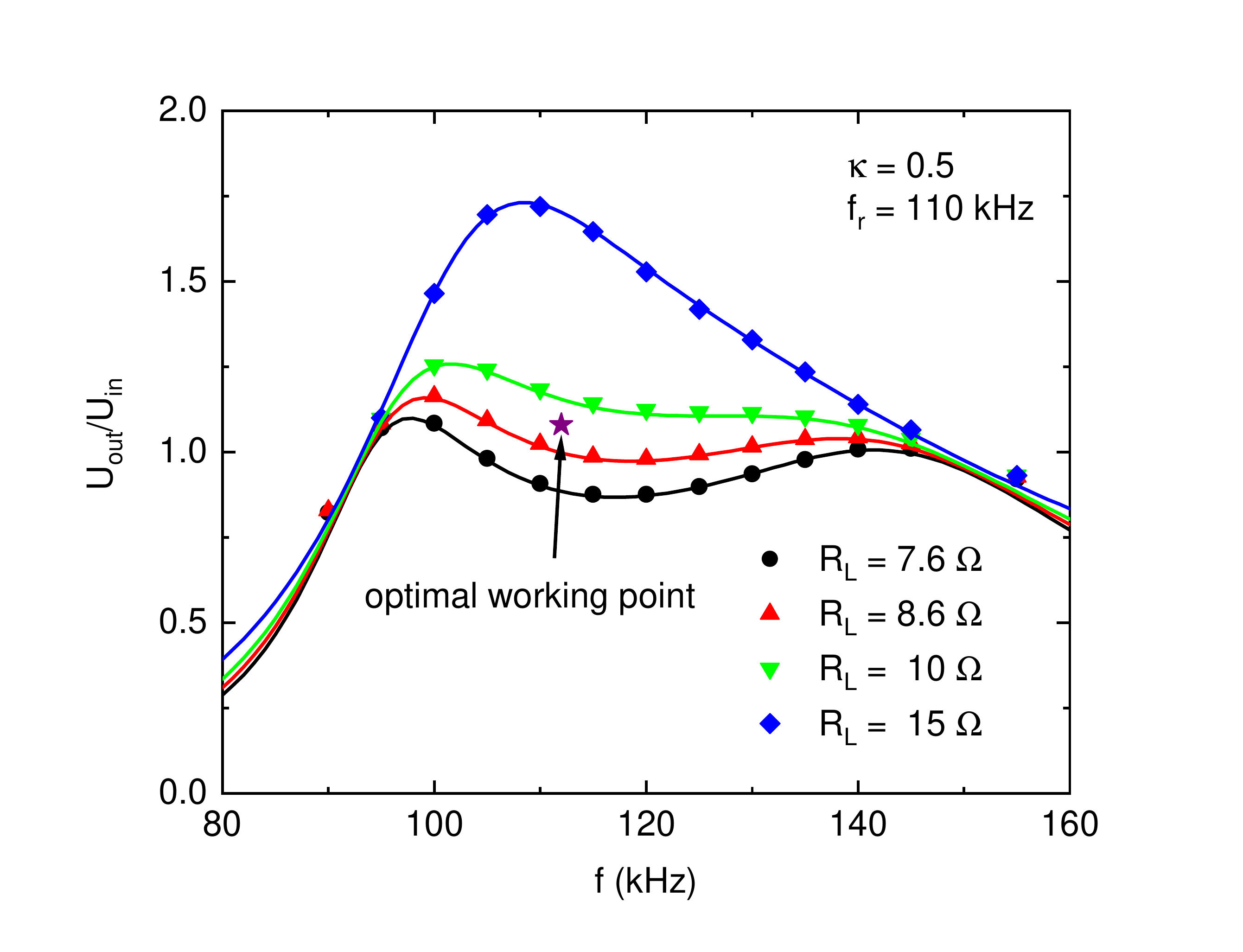}
	\caption{Measured (symbols) and calculated \cite{Bosshard.2015} (solid lines) voltage gain of the WPT system from the source to the load. The resonance frequencies of the transmitter and the receiver circuits are fixed to $f_\mathrm{r}=\SI{110}{kHz}$. The drive tone frequency is swept from $\SI{85}{kHz}$ to $\SI{155}{kHz}$ to find the optimal working point (purple asterisk). Comparison of different load values shows the transition from the under-coupled into the over-coupled regime. The superconducting system behaves as expected for a conventional system.}
	\label{fig:freq-sweep}
\end{figure}

\begin{figure}[!t]
	\centering
	\includegraphics[width=0.45\textwidth]{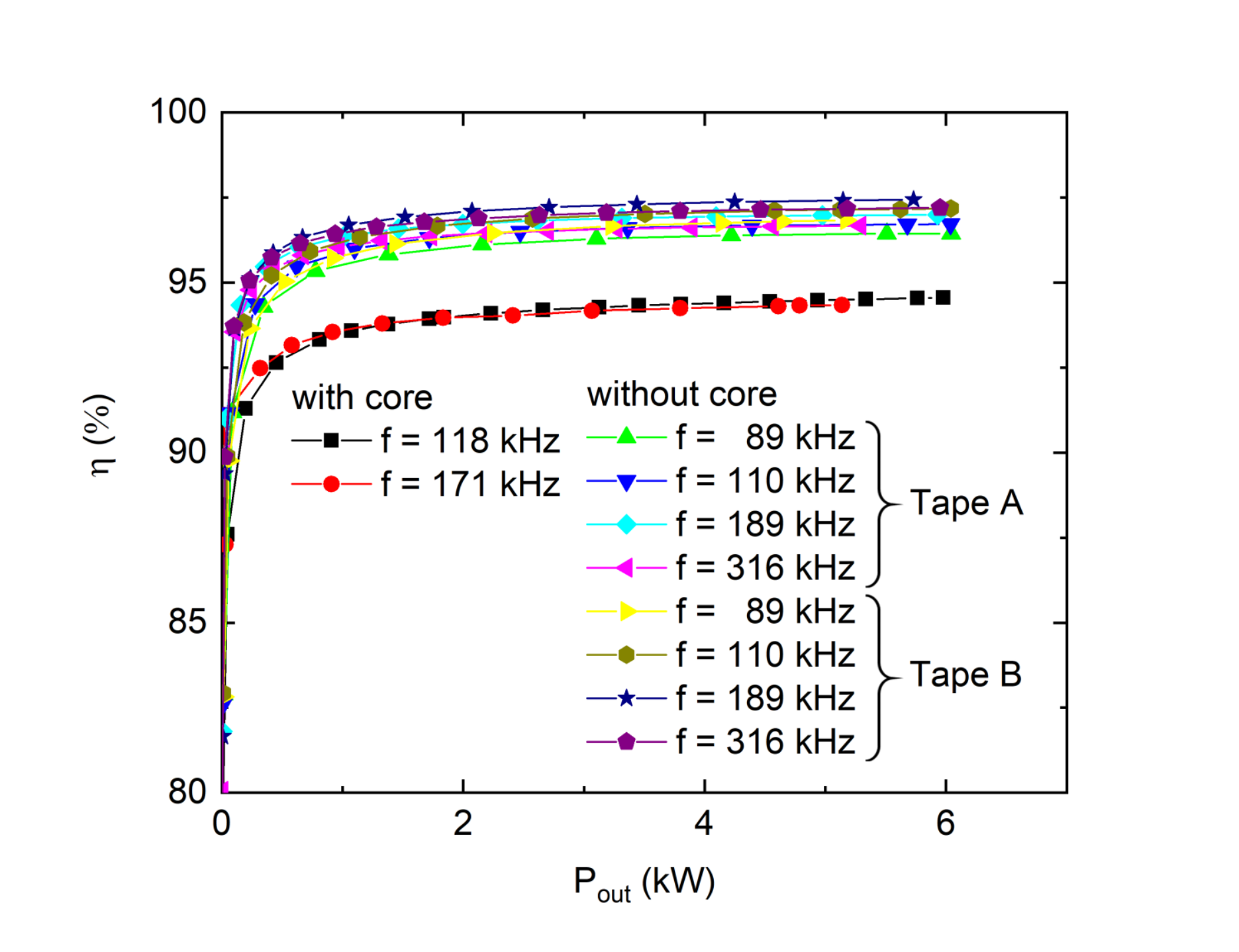}
	\caption{Full system DC-to-DC efficiency as a function of the output power. Measurements with and without magnetic core are compared to each other. Without the core, the efficiency is limited by eddy current losses, a frequency dependence becomes visible and the stabilized coils outperform their non-stabilized counterparts. An efficiency of $\eta=\SI{97.44}{\percent}$ has been achieved at an output power of $\SI{6}{kW}$. Higher power levels were not accessible with the available setup.}
	\label{fig:DC-to-DC-efficiency}
\end{figure}

\begin{figure}[!t]
	\centering
	\includegraphics[width=0.45\textwidth]{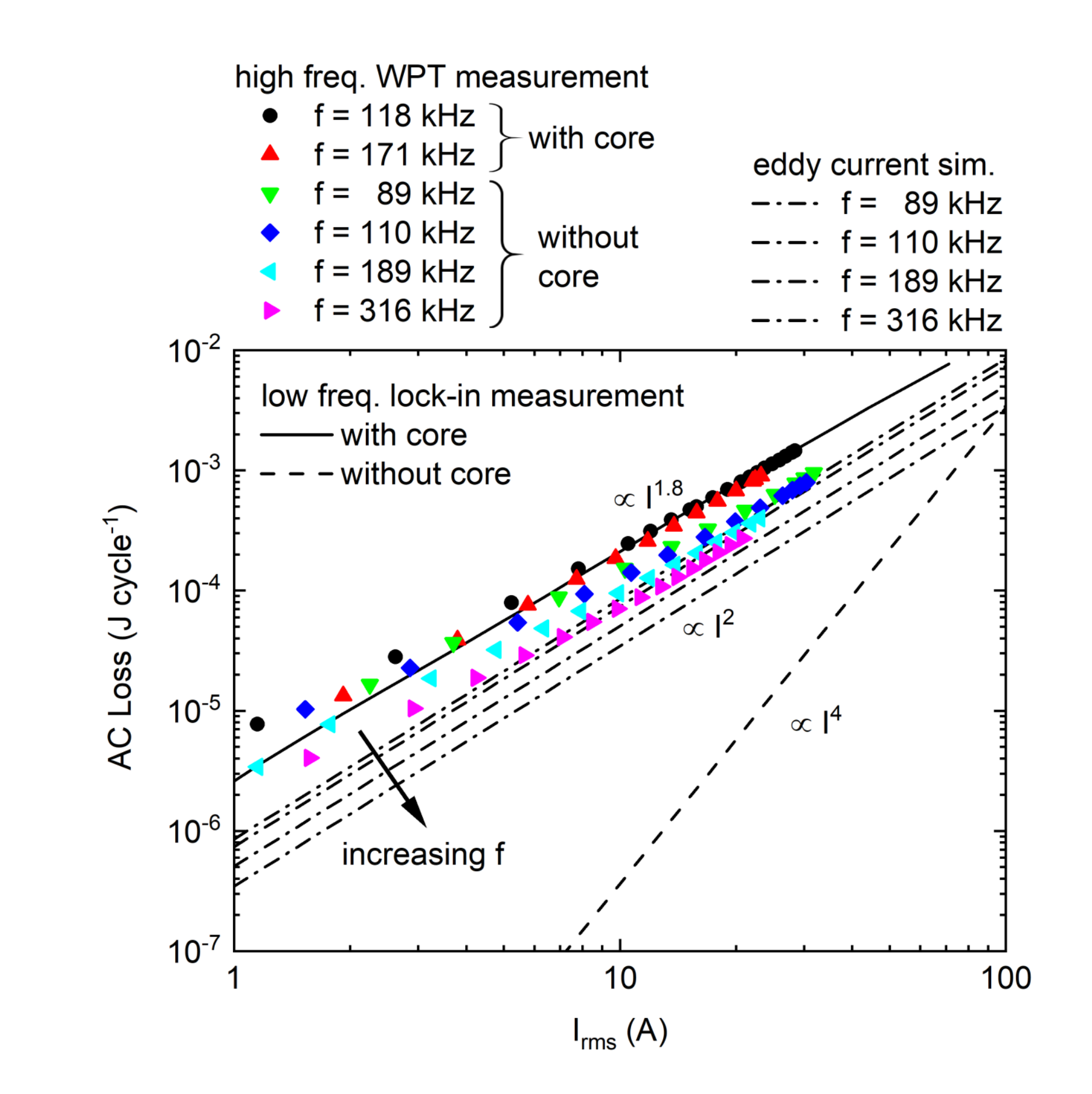}
	\caption{Measured AC loss of the stabilized HTS coil (Tape B), with and without magnetic core, in the high frequency regime. The results are extracted from DC-to-DC loss measurements of the WPT system, and are compared to the low frequency lock-in measurements. The magnetic core loss shows perfect agreement between both methods. Without magnetic core, the eddy current losses become observable. Quantitative agreement to the numerical eddy current simulations can be found for large current amplitudes. At low current amplitudes and at frequencies beyond $\SI{200}{kHz}$, parasitic losses in the power electronics become significant and distort the measurement slightly.}
	\label{fig:ac-loss-high-freq}
\end{figure}

Both fabricated coil pairs were used to characterize the full WPT system. Firstly, the coupling constant between the transmitter and the receiver coil  was measured as a function of the transfer distance. The results for both cases, with and without magnetic core, are shown in Fig. \ref{fig:coupling}.
The influence of the magnetic core on the coupling strength is small and the coil pair can therefore be used at transfer distances of up to $\SI{10}{cm}$, if a small coupling constant of $\kappa\approx 0.1$ is acceptable in the system. For simplicity and comparability, all of the following WPT measurements were performed at a fixed coupling strength of $\kappa\approx 0.5$.\\
The loss measurements on single coils, presented above, have shown that the current amplitudes through the HTS coils must remain low to benefit from high Q-factors. Therefore, the WPT system was operated at high voltages and low currents. A reasonable choice of target parameters for an output power of $\SI{11}{kW}$ would be $U\approx \SI{300}{V}$ and $I\approx \SI{36}{A}$, which fixes the load resistance at the receiver side to \mbox{$R_\mathrm{L}\approx\SI{8.5}{\ohm}$}. According to Eq. (\ref{eq:impedance-matching}), the required operating frequency to reach an efficient power transfer, with the available coil inductance of $L=\SI{20}{\mu \henry}$, is in the order of $\SI{100}{kHz}$. Clearly this frequency is in the regime where eddy currents dominate the total loss of the coils. For application in the low frequency regime a higher coil inductance would be required. Nevertheless, the overall performance and especially the frequency dependence in the high frequency regime was studied.\\
By tolerating small deviations from the nominal current and voltage values, the system was operated in a frequency range from $f_\mathrm{r}=\SI{89}{kHz}$ to $f_\mathrm{r}=\SI{316}{kHz}$. For each measurement, conduction cooled power capacitors (CSM Nano and CSP 120/200) from the manufacturer Celem were used, to tune the resonance frequency of the LC circuit to the desired value. The system was then driven at the optimal working point, where $R_\mathrm{L}$ satisfies Eq. (\ref{eq:impedance-matching}) and the drive frequency is closely above $f_\mathrm{r}$. A photograph of the experimental setup including all measurement devices that are required for the full system characterization is depicted in Fig. \ref{fig:full-system}.\\
In order to proof that the system, with superconducting coils on both sides, can be described with the analytical model of conventional systems and that the design considerations are correct, the voltage gain from source to load was measured as a function of the drive tone frequency at a fixed resonance frequency of \mbox{$f_\mathrm{r}=\SI{110}{kHz}$}. The expected load value which separates the over- and the under-coupled regimes is $R_\mathrm{L}=\SI{9.2}{\ohm}$. In Fig. \ref{fig:freq-sweep}, the measured voltage response is shown, at load values of $\SI{7.6}{\ohm}$, $\SI{8.6}{\ohm}$, $\SI{10}{\ohm}$ and $\SI{15}{\ohm}$, together with the analytical calculation according to Ref. \cite{Bosshard.2015} (solid lines).
The optimal working point of this configuration is indicated with a purple asterisk. The system  behavior agrees exactly with the expectation for a conventional system and the transition from the under-coupled regime, where a frequency splitting is present, into the over-coupled regime, where a voltage gain around $f_\mathrm{r}$ appears, can be observed.\\
In Fig. \ref{fig:DC-to-DC-efficiency}, the DC-to-DC efficiency of the full system is presented as a function of the output power at different operating conditions. Measurements with- and without magnetic core are compared to each other. With the core, the efficiency is limited to $\eta=\SI{95}{\percent}$ and the system performance is frequency independent. Without the core, $\eta$ is limited by eddy current losses and becomes frequency dependent. The efficiency increases with increasing frequency and the stabilized coils outperform their non-stabilized counterparts. A maximum efficiency of $\eta=\SI{97.44}{\percent}$ has been achieved at an output power of $\SI{6}{kW}$. Higher power levels were not accessible with the available setup, as the system was either limited by the voltage of the DC source, or the current amplitudes reached such high values that the AC losses exceeded the cooling power of the liquid nitrogen bath and the HTS coils quenched. A quench at the maximum output power of $\SI{6}{kW}$ did however not damage the coils. The relatively low coil currents can easily be taken over by the metal layers of the tapes. Even the non-stabilized coils were driven several times into a thermal quench at a coil current of $I_\mathrm{rms}\approx \SI{40}{A}$, without taking damage.\\
The results, presented in Fig. \ref{fig:DC-to-DC-efficiency}, have been further used to study the AC loss of the coils in the high frequency regime. With the assumption that the losses in the power electronics are small, i.e. the measured power loss of the complete WPT system is dissipated mostly in the transmission coils, the DC-to-DC measurements can be used to extract the AC loss of both coils. If the current amplitudes in both coils are similar, then the loss per cycle per coil is given by $P_\mathrm{loss,tot}/(2f)$. In reality, a small fraction of the power loss is, of course, dissipated in the room temperature electronics. In order not to further extend the scope of this paper, a detailed discussion of the power losses in the MOSFETs, the resonant capacitors and the Schottky diodes is omitted. Nevertheless, our measurement provides an upper limit for the high-frequency AC loss in the HTS coil, which is valid in any case. The fact that the power loss in the setup is measured in DC values, enables high precision measurements of the AC loss, at frequencies which are not accessible by typical lock-in techniques. The power loss per coil, as measured with the WPT setup, is presented in Fig. \ref{fig:ac-loss-high-freq}.
The data is compared to the results of the low frequency lock-in measurement and to numerical eddy current simulations. The extracted AC loss data was further used to calculate the quality factors of the coils under high frequency conditions with Eq. (\ref{eq:quality-factor}). The results are shown in Fig. \ref{fig:q-factor}. In both, Fig. \ref{fig:q-factor} and Fig. \ref{fig:ac-loss-high-freq}, good agreement between the low frequency lock-in measurements and the high frequency WPT experiments is found for the case that the magnetic core is applied. This proofs that our high frequency loss measurement gives correct results and allows us to study the eddy current losses. Without the core, the power losses at high frequencies are strongly enhanced, compared to the low frequency lock-in measurement. The frequency dependence of the eddy current loss is observable, the measured power loss scales with $I^2$ and agrees for large current amplitudes with the numerical simulation. The small deviation between measurement and simulation at low current amplitudes and at frequencies beyond $\SI{200}{kHz}$ can be explained with parasitic losses in the power electronics.\\
Analysis of the same experiments with the non-stabilized coil pairs has also given consistent results. The eddy current loss was increased, and the frequency dependence was reduced, compared to the stabilized coil pair. As depicted in Fig. \ref{fig:eddy-current-loss}, the numerical simulation suggests that the thin silver layer of tape A generates more eddy current loss than the copper layer of tape B at frequencies around $\SI{100}{kHz}$. Further, the eddy current loss of the silver layer has a maximum in the studied frequency range, which explains the reduction of the frequency dependence. The only options to get rid of eddy current losses in WPT systems are either, to operate the HTS coils in the low frequency regime $f<f_\mathrm{tr}$, or to use completely non-metallic tapes. Here the development of sapphire based substrates \cite{Tixador.2019} could possibly bring benefits in the future.

\section{Conclusion and Outlook}
In this paper a fully functional superconducting WPT system, consisting of optimized single pancake HTS coils on the transmitter and on the receiver side, was presented. For the first time, the successful power transfer of significant power levels was demonstrated. At an output power of $\SI{6}{kW}$, a DC-to-DC efficiency of $\SI{97.44}{\percent}$ has been achieved. The area- and weight-related power densities of the transmission coils are $\SI{1.59}{kW/dm^2}$ and $\SI{5.71}{kW/kg}$. With a thickness of only $\SI{2}{cm}$, the HTS coil is very thin, what results in a high volumetric power density of $\SI{7.95}{kw/dm^3}$. All three values exceed the power densities of conventional state-of-the-art transmission coils \cite{Bosshard.2015,BosshardPhD}. \\
These results have been achieved by introducing a coil design with a distributed HTS winding, where the individual turns are separated by an inter-tape spacing. The superconducting hysteresis loss and the metallic eddy current losses for such a winding were calculated and optimized. The results have been validated by a standard lock-in technique in the low frequency regime and by WPT experiments in the high frequency regime. It was found that an inter-tape spacing of $d=w/2$ is a good compromise to keep both, the coil size and the AC losses, small. The measured quality-factors of the fabricated coil prototypes exceed values of $10000$, making them a very promising candidate for further studies. Despite the fact that all WPT experiments have been performed in the high frequency regime, where eddy currents dominate the loss of the coil, the presented system showed an outstanding performance. It is expected that the efficiency and the accessible power levels could be further increased by reducing the operating frequency below $f_\mathrm{tr}$. The authors plan to apply the derived design guidelines to build a new pair of transmission coils with sufficiently large inductance. This will allow to operate the system at lower frequencies, where the lock-in measurements already have demonstrated the big potential of the presented coils.\\
Regarding future applications of this system in electrical machines, the efficient cooling of the coils without losing the argument of high power density represents a key challenge. Obviously, the cryostat design needs to be completely non-metallic and the required regular access to liquid nitrogen could be an obstacle for autonomous machines.\\


%

\appendices




\ifCLASSOPTIONcaptionsoff
  \newpage
\fi



\bibliographystyle{IEEEtran}
%
\bibliography{phd_lit}

\begin{thebibliography}{10}
\providecommand{\url}[1]{#1}
\csname url@samestyle\endcsname
\providecommand{\newblock}{\relax}
\providecommand{\bibinfo}[2]{#2}
\providecommand{\BIBentrySTDinterwordspacing}{\spaceskip=0pt\relax}
\providecommand{\BIBentryALTinterwordstretchfactor}{4}
\providecommand{\BIBentryALTinterwordspacing}{\spaceskip=\fontdimen2\font plus
\BIBentryALTinterwordstretchfactor\fontdimen3\font minus
  \fontdimen4\font\relax}
\providecommand{\BIBforeignlanguage}[2]{{%
\expandafter\ifx\csname l@#1\endcsname\relax
\typeout{** WARNING: IEEEtran.bst: No hyphenation pattern has been}%
\typeout{** loaded for the language `#1'. Using the pattern for}%
\typeout{** the default language instead.}%
\else
\language=\csname l@#1\endcsname
\fi
#2}}
\providecommand{\BIBdecl}{\relax}
\BIBdecl

\bibitem{MomentumDynamics.2018}
\BIBentryALTinterwordspacing
``Martha's vineyard buses get wirelessly charged up with 200-kw system,'' 2018.
  [Online]. Available: \url{https://momentumdynamics.com}
\BIBentrySTDinterwordspacing

\bibitem{IPTTechnology.2019}
\BIBentryALTinterwordspacing
``Wireless charging for a smooth and safe power transfer from shore to the
  ferry,'' 2019. [Online]. Available: \url{https://ipt-technology.com}
\BIBentrySTDinterwordspacing

\bibitem{Wave.2019}
\BIBentryALTinterwordspacing
``Wave drayage electrification project,'' 2019. [Online]. Available:
  \url{https://waveipt.com}
\BIBentrySTDinterwordspacing

\bibitem{Bosshard.2015}
R.~Bosshard, J.~W. Kolar, J.~Muhlethaler, I.~Stevanovic, B.~Wunsch, and
  F.~Canales, ``Modeling and $\eta$-$\alpha$-pareto optimization of inductive
  power transfer coils for electric vehicles,'' \emph{IEEE Journal of Emerging
  and Selected Topics in Power Electronics}, vol.~3, no.~1, pp. 50--64, 2015.

\bibitem{XiaoYuanChenJianXunJin.2011}
\BIBentryALTinterwordspacing
{Xiao Yuan Chen, Jian Xun Jin}, \emph{Resonant Circuit and Magnetic Field
  Analysis of Superconducting Contactless Power Transfer: 14 - 16 Dec. 2011,
  Sydney, Australia}.\hskip 1em plus 0.5em minus 0.4em\relax Piscataway, NJ:
  IEEE, 2011. [Online]. Available:
  \url{http://ieeexplore.ieee.org/servlet/opac?punumber=6135857}
\BIBentrySTDinterwordspacing

\bibitem{Kim.2012}
D.~W. Kim, Y.~D. Chung, H.~K. Kang, Y.~S. Yoon, and T.~K. Ko, ``Characteristics
  of contactless power transfer for hts coil based on electromagnetic resonance
  coupling,'' \emph{IEEE Transactions on Applied Superconductivity}, vol.~22,
  no.~3, p. 5400604, 2012.

\bibitem{Zuo.2015}
W.~Zuo, S.~Wang, Y.~Liao, and Y.~Xu, ``Investigation of efficiency and load
  characteristics of superconducting wireless power transfer system,''
  \emph{IEEE Transactions on Applied Superconductivity}, vol.~25, no.~3, pp.
  1--6, 2015.

\bibitem{Chen.2016}
X.~Y. Chen, J.~X. Jin, L.~H. Zheng, and Z.~H. Wu, ``A rotary-type contactless
  power transfer system using hts primary,'' \emph{IEEE Transactions on Applied
  Superconductivity}, vol.~26, no.~7, pp. 1--6, 2016.

\bibitem{Narayanamoorthi.2019}
R.~Narayanamoorthi and A.~V. Juliet, ``Capacitor-less high-strength resonant
  wireless power transfer using open bifilar spiral coil,'' \emph{IEEE
  Transactions on Applied Superconductivity}, vol.~29, no.~1, pp. 1--8, 2019.

\bibitem{Li.2019}
W.~Li, T.~W. Ching, C.~Jiang, T.~Wang, and {Le Sun}, ``Quantitative comparison
  of wireless power transfer using hts and copper coils,'' \emph{IEEE
  Transactions on Applied Superconductivity}, vol.~29, no.~5, pp. 1--6, 2019.

\bibitem{Zhang.2012}
M.~Zhang, J.-H. Kim, S.~Pamidi, M.~Chudy, W.~Yuan, and T.~A. Coombs, ``Study of
  second generation, high-temperature superconducting coils: Determination of
  critical current,'' \emph{Journal of Applied Physics}, vol. 111, no.~8, p.
  083902, 2012.

\bibitem{Yu.2015}
H.~Yu, G.~Zhang, L.~Jing, Q.~Liu, W.~Yuan, Z.~Liu, and X.~Feng, ``Wireless
  power transfer with hts transmitting and relaying coils,'' \emph{IEEE
  Transactions on Applied Superconductivity}, vol.~25, no.~3, pp. 1--5, 2015.

\bibitem{Kang.2014}
H.~K. Kang, Y.~D. Chung, and S.~W. Yim, ``Conceptual design of contactless
  power transfer into hts receiver coil using normal conducting resonance
  antenna,'' \emph{Cryogenics}, vol.~63, pp. 209--214, 2014.

\bibitem{Inoue.2017b}
R.~Inoue, D.~Miyagi, M.~Tsuda, and H.~Matsuki, ``High-efficiency transmission
  of a wireless power transmission system for low-frequency using rebco
  double-pancake coils,'' \emph{IEEE Transactions on Applied
  Superconductivity}, vol.~27, no.~1, pp. 1--6, 2017.

\bibitem{Pardo.2012b}
E.~Pardo, J.~{\v{S}}ouc, and J.~Kov{\'a}{\v{c}}, ``Ac loss in rebco pancake
  coils and stacks of them: modelling and measurement,'' \emph{Superconductor
  Science and Technology}, vol.~25, no.~3, p. 035003, 2012.

\bibitem{Shen.2018}
B.~Shen, C.~Li, J.~Geng, X.~Zhang, J.~Gawith, J.~Ma, Y.~Liu, F.~Grilli, and
  T.~A. Coombs, ``Power dissipation in hts coated conductor coils under the
  simultaneous action of ac and dc currents and fields,'' \emph{Superconductor
  Science and Technology}, vol.~31, no.~7, p. 075005, 2018.

\bibitem{Shen.2017c}
B.~Shen, J.~Li, J.~Geng, L.~Fu, X.~Zhang, C.~Li, H.~Zhang, Q.~Dong, J.~Ma, and
  T.~A. Coombs, ``Investigation and comparison of ac losses on stabilizer-free
  and copper stabilizer hts tapes,'' \emph{Physica C: Superconductivity and its
  Applications}, vol. 541, pp. 40--44, 2017.

\bibitem{Inoue.2018}
R.~Inoue, D.~Miyagi, M.~Tsuda, and H.~i. Matsuki, ``Magnetization loss
  characteristics of a gdbco tape in khz frequency band,'' \emph{IEEE
  Transactions on Applied Superconductivity}, vol.~28, no.~4, pp. 1--5, 2018.

\bibitem{Zhou.2019c}
P.~Zhou, G.~Ma, and L.~Qu{\'e}val, ``Transition frequency of transport ac
  losses in high temperature superconducting coated conductors,'' \emph{Journal
  of Applied Physics}, vol. 126, no.~6, p. 063901, 2019.

\bibitem{Wang.2004}
C.-S. Wang, G.~A. Covic, and O.~H. Stielau, ``Power transfer capability and
  bifurcation phenomena of loosely coupled inductive power transfer systems,''
  \emph{IEEE Transactions on Industrial Electronics}, vol.~51, no.~1, pp.
  148--157, 2004.

\bibitem{Budhia.2011}
M.~Budhia, G.~A. Covic, and J.~T. Boys, ``Design and optimization of circular
  magnetic structures for lumped inductive power transfer systems,'' \emph{IEEE
  Transactions on Power Electronics}, vol.~26, no.~11, pp. 3096--3108, 2011.

\bibitem{Roshen.1988}
W.~A. Roshen and D.~E. Turcotte, ``Planar inductors on magnetic substrates,''
  \emph{IEEE Transactions on Magnetics}, vol.~24, no.~6, pp. 3213--3216, 1988.

\bibitem{Grilli.2014b}
F.~Grilli, V.~M.~R. Zermeno, E.~Pardo, M.~Vojenciak, J.~Brand, A.~Kario, and
  W.~Goldacker, ``Self-field effects and ac losses in pancake coils assembled
  from coated conductor roebel cables,'' \emph{IEEE Transactions on Applied
  Superconductivity}, vol.~24, no.~3, pp. 1--5, 2014.

\bibitem{Muller.1997}
K.-H. M{\"u}ller, ``Ac power losses in flexible thick-film superconducting
  tapes,'' \emph{Physica C: Superconductivity}, vol. 281, no.~1, pp. 1--10,
  1997.

\bibitem{Grilli.2014}
F.~Grilli, E.~Pardo, A.~Stenvall, D.~N. Nguyen, W.~Yuan, and F.~Gomory,
  ``Computation of losses in hts under the action of varying magnetic fields
  and currents,'' \emph{IEEE Transactions on Applied Superconductivity},
  vol.~24, no.~1, pp. 78--110, 2014.

\bibitem{Muller.1999}
K.-H. M{\"u}ller, ``Ac losses in stacks and arrays of ybco/hastelloy and
  monofilamentary bi-2223/ag tapes,'' \emph{Physica C: Superconductivity}, vol.
  312, no. 1-2, pp. 149--167, 1999.

\bibitem{Schonborg.2001}
N.~Sch{\"o}nborg, ``Hysteresis losses in a thin high-temperature superconductor
  strip exposed to ac transport currents and magnetic fields,'' \emph{Journal
  of Applied Physics}, vol.~90, no.~6, pp. 2930--2933, 2001.

\bibitem{Lu.2008}
J.~Lu, E.~S. Choi, and H.~D. Zhou, ``Physical properties of hastelloy c-276 at
  cryogenic temperatures,'' \emph{Journal of Applied Physics}, vol. 103, no.~6,
  p. 064908, 2008.

\bibitem{Ekin.2007}
J.~W. Ekin, \emph{Experimental techniques for low-temperature measurements:
  Cryostat design, material properties, and superconductor critical-current
  testing}, reprinted~ed.\hskip 1em plus 0.5em minus 0.4em\relax Oxford:
  {Oxford Univ. Press}, 2007.

\bibitem{Clem.2007}
J.~R. Clem, J.~H. Claassen, and Y.~Mawatari, ``Ac losses in a finite z stack
  using an anisotropic homogeneous-medium approximation,'' \emph{Superconductor
  Science and Technology}, vol.~20, no.~12, pp. 1130--1139, 2007.

\bibitem{Safran.2017}
S.~Safran, J.~Souc, and F.~G{\"o}m{\"o}ry, ``Ac loss characterization of single
  pancake bscco coils by measured different methods,'' \emph{Physica C:
  Superconductivity and its Applications}, vol. 541, pp. 45--49, 2017.

\bibitem{Norris1970}
W.~T. Norris, ``Calculation of hysteresis losses in hard superconductors
  carrying ac: isolated conductors and edges of thin sheets,'' \emph{Journal of
  Physics D: Applied Physics}, vol.~3, no. 489, 1970.

\bibitem{Gomory.2013}
F.~Gomory, J.~Souc, E.~Pardo, E.~Seiler, M.~Soloviov, L.~Frolek, M.~Skarba,
  P.~Konopka, M.~Pekarcikova, and J.~Janovec, ``Ac loss in pancake coil made
  from 12 mm wide rebco tape,'' \emph{IEEE Transactions on Applied
  Superconductivity}, vol.~23, no.~3, p. 5900406, 2013.

\bibitem{Souc.2005}
J.~{\v{S}}ouc, F.~G{\"o}m{\"o}ry, and M.~Vojen{\v{c}}iak, ``Calibration free
  method for measurement of the ac magnetization loss,'' \emph{Superconductor
  Science and Technology}, vol.~18, no.~5, pp. 592--595, 2005.

\bibitem{Claassen.2005}
J.~H. Claassen, ``Inductor design for cryogenic power electronics,'' \emph{IEEE
  Transactions on Appiled Superconductivity}, vol.~15, no.~2, pp. 2385--2388,
  2005.

\bibitem{Tixador.2019}
P.~Tixador, M.~Bauer, C.-E. Bruzek, A.~Calleja, G.~Deutscher, B.~Dutoit,
  F.~Gomory, L.~Martini, M.~Noe, X.~Obradors, M.~Pekarcikova, and F.~Sirois,
  ``Status of the european union project fastgrid,'' \emph{IEEE Transactions on
  Applied Superconductivity}, vol.~29, no.~5, pp. 1--5, 2019.

\bibitem{BosshardPhD}
R.~Bosshard, ``Multi-objective optimization of inductive power transfer systems
  for ev charging,'' PhD Thesis, {ETH Zurich}, 2015.

\end{thebibliography}

%








\end{document}